\documentclass[%
reprint,
superscriptaddress,
showpacs,
 amsmath,amssymb,
 aps,
]{revtex4-1}

\usepackage[percent]{overpic}
\usepackage{tikz,siunitx,mwe}
\usepackage{graphicx,color,layout}
\usepackage{dcolumn}
\usepackage{bm}
\usepackage[mathlines]{lineno}

\begin{document}


\title{Exclusion and Verification of Remote Nuclear Reactors \\with a 1-Kiloton Gd-Doped Water Detector } 


\author{O. A. Akindele}
\email[]{Corresponding author; Email: akindele1@llnl.gov}
\affiliation{Lawrence Livermore National Laboratory, Livermore, California 94550, USA}

\author{A. Bernstein}
\author{ M. Bergevin} 
\author{S. A. Dazeley}
\author{F. Sutanto}
\author{A. Mullen}
\author{J. Hecla}
\affiliation{Lawrence Livermore National Laboratory, Livermore, California 94550, USA}
\date{\today}
\begin{abstract}

To date, antineutrino experiments built for the purpose of demonstrating a nonproliferation capability have typically employed organic scintillators, were situated as close to the core as possible - 
typically a few meters to tens of meters distant and have not exceeded a few tons in size. 
One problem with this approach is that proximity to the reactor core require accommodation by the host facility. Water Cherenkov detectors located offsite, at distances of a few kilometers or greater, may facilitate non-intrusive monitoring and verification of reactor activities over a large area. 
As the standoff distance increases, the detector target mass must scale accordingly.  
This article quantifies the degree to which a kiloton-scale gadolinium-doped water-Cherenkov  detector can exclude the existence of undeclared reactors within a specified distance, and remotely detect the presence of a hidden reactor in the presence of declared reactors, by verifying the operational power and standoff distance using a Feldman-Cousins based likelihood analysis. A 1-kton scale (fiducial) water Cherenkov detector can exclude gigawatt-scale nuclear reactors up to tens of kilometers within a year. When attempting to identify the specific range and power of a reactor, the detector energy resolution was not sufficient to delineate between the two.


\end{abstract}

\maketitle


\section{\label{sec:intro}Introduction}                                                  


Antineutrino monitoring has been proposed for various nonproliferation applications and reactor fuel compositions \cite{bowden2007experimental,christensen2015antineutrino,bernstein2018reactors,akindele2016antineutrino,bernstein2020colloquium,hayes2012antineutrino}. 
All of these prior studies used ton-scale scintillator detectors in close proximity (tens of meters or less) to the reactor core. 
While short-distance monitoring with scintillator detectors is less intrusive to reactor operations compared to other verification methods, these detectors still require on-site accommodation. For example, space and power must be provided for the equipment, the materials used must be compliant with facility regulations, and any maintenance on the detector will require access by a verification body. 
Greater standoff distances could further reduce the intrusiveness of the method by permitting deployment outside the reactor operator's facility grounds.  However, to effectively probe a larger exclusion area requires a larger target volume.
Ton-scale near-field detectors still require on-site compliance. Space and power must be provided  for the equipment, the materials used must be shown not to affect facility operations, and any maintenance on the detector will require access by the verification body.

Antineutrinos are weakly interacting and can be detected at long distances from their source of origin. This raises the question: can antineutrino detection be used in the mid-field, which we define here to be approximately 10 to 100 km, to monitor the operation or presence of nuclear reactors? In this article, we evaluate the ability to detect antineutrinos at a hypothetical far-field deployment in the presence of other reactors producing a high antineutrino background.


\section{\label{sec:WATCHMAN}The  baseline detector design}
\begin{figure*}[ht]
\includegraphics[width=0.9\linewidth]
{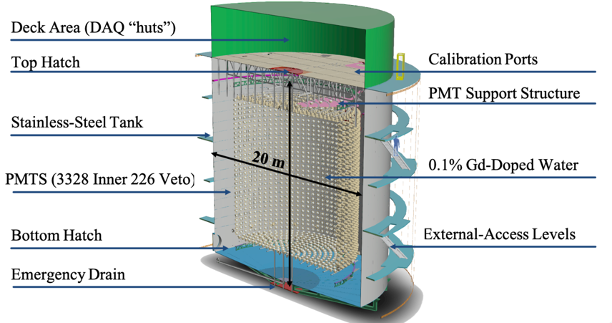}

\caption{ An illustration of the Gd-H$_2$O baseline detector design, with a vertical cutaway through the center. The width and height of the cylindrical tank is 20-meters. In total, the detector will house 3,554 Photo-multiplier Tubes (PMTs), 3,328 as part of the inner detector (20 $\%$ photocoverage) facing the fiducial volume, and 226 veto PMTs facing the tank (2 $\%$ photocoverage)to reject cosmogenic events from within the veto volume. Other components of the tank include: a top and bottom hatch, calibration ports, a top deck for data acquisition (DAQ) electronics, and external access locations.}
\label{fig_WM_CAD}.
\end{figure*}

In this study the hypothetical deployment is located in the Boulby Underground Laboratory on the eastern side of the United Kingdom,  an operating potash/polyhalite mine \cite{robinson2003measurements}. The mine rock is low in uranium, thorium, and radon compared to many other underground facilities. 
The detector is modeled to be $1.1 \ \mathrm{km}$ (2.86 km.w.e.) underground, and approximately 26 km from the Hartlepool Reactor Complex. The Complex houses two 1.5 GW\textsubscript{th} Advanced Gas Reactors (AGRs) yielding 3 GW\textsubscript{th} capacity. 
The reactor antineutrino background at the Boulby mine is relatively high 
due to the presence of a large number of other operating reactors in the UK and Western Europe. 
For specific use cases, design variations may be employed to maximize the sensitivity of the detector, such as increased photo-coverage, large target volumes, or multiple detectors.

This article presents the sensitivity of a gadolinium-doped water (Gd-H$_2$O) detector, using the  WATCHMAN collaboration's 2019 baseline Gd-H$_2$O detector design \cite{bernstein2019ait}. We refer to this design as \lq the Gd-H$_2$O baseline design' or the \lq Gd-H$_2$O detector'         throughout this article. Sensitivity estimates are provided for exclusion of the existence of undeclared reactors over a specified radial distance, and for determining the presence of a hidden reactor near a declared reactor facility.  The Boulby Underground Laboratory site is used to provide a concrete example of sensitivity in a well-studied background environment \cite{robinson2003measurements}.

The detection medium is contained in a cylindrical stainless-steel tank with a 20-meter height and diameter. The tank is filled with approximately 6 kilotons of ultra-pure water mixed with gadolinium sulfate, for a total loading of 0.1$\%$ gadolinium by weight. 
The detector has two optically separated regions, the muon veto region and the inner detector. Events occurring in the 3.3 m thick outer veto region 
are read out by 226 PMTs, while events in the inner detector are read out by 3,328 PMTs. 
This equates to a 20$\%$ photo-coverage for the inner detector and a 2$\%$ photo-coverage for the veto detector.
To reduce backgrounds caused by radiation from the PMTs, only events which reconstruct within the central $\sim$1 kton fiducial volume will be considered as candidate antineutrinos. An illustration of the detector is shown in Figure \ref{fig_WM_CAD}.

Gadolinium-doped water is an ideal medium for far-field antineutrino monitoring due to its chemical simplicity, low cost, and good optical transparency\cite{,beacom2004antineutrino,bernstein2001assessment}. The long attenuation length of the Gd-doped water
allows the medium to be scaled to larger volumes if required, based on the particular needs of real world applications \cite{marti2020evaluation,renshaw2012research}. 
Water is a low toxicity  medium that has been deployed in many other neutrino detectors such as Super Kamiokande\cite{SuperK}, and SNO\cite{SNO}. 
The addition of the capture agent gadolinium allows for an enhanced neutron capture signal, favorable for experiments focused primarily on detecting reactor antineutrinos through inverse beta decay (IBD). 

In the IBD process, antineutrinos interact with quasi-free protons in the water, producing a positron-neutron pair in the final state:
\begin{equation}
    \overline{\nu}_{e}+p\,\longrightarrow e^{+}+n.
\end{equation}
The positron is detected as a prompt signal through the Cherenkov light emitted when its velocity exceeds the speed of light in the water.
This is a multiple threshold reaction in which the $\overline{\nu}_e$ energy must exceed 1.8 MeV to generate an IBD reaction, and the resulting positron kinetic energy must exceed $\sim$253 keV in water to generate Cherenkov light.  
The neutron produced through IBD will elastically scatter off hydrogen in the detector until thermalization, after which it can capture on either a gadolinium or hydrogen nucleus. 
For gadolinium loading at 0.1\%, the average capture time is $\sim$30 $\mu$s after the prompt positron event \cite{beacom2004antineutrino}. Neutron capture on hydrogen account for 9$\%$ of captures and will result in a single 2.2-MeV gamma ray, while captures on isotopes of gadolinium accounts for 91$\%$ of captures and will result in a cascade of gamma rays summing to $\sim$8 MeV. 
The gamma rays from neutron capture will Compton scatter, and the Compton electrons may emit Cherenkov light. 
Due to the threshold required for Cherenkov emission, not all of the scattered electrons from gamma rays released from neutron capture on Gd will contribute to the signal, but the neutron capture on Gd still produces a distinctly bright signature.

\begin{figure}[]

\includegraphics[width=0.97\linewidth]{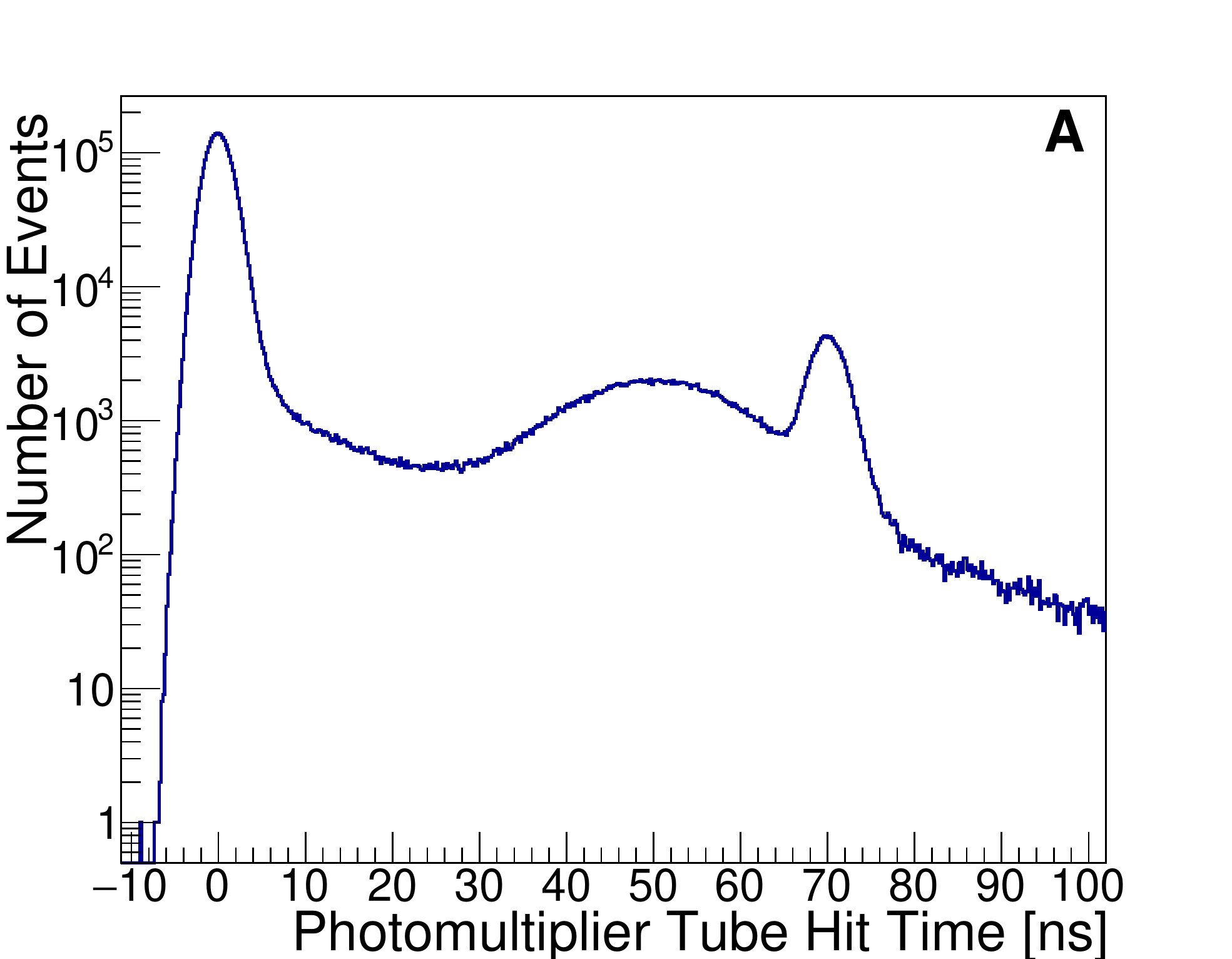} 
\includegraphics[width=0.97\linewidth]{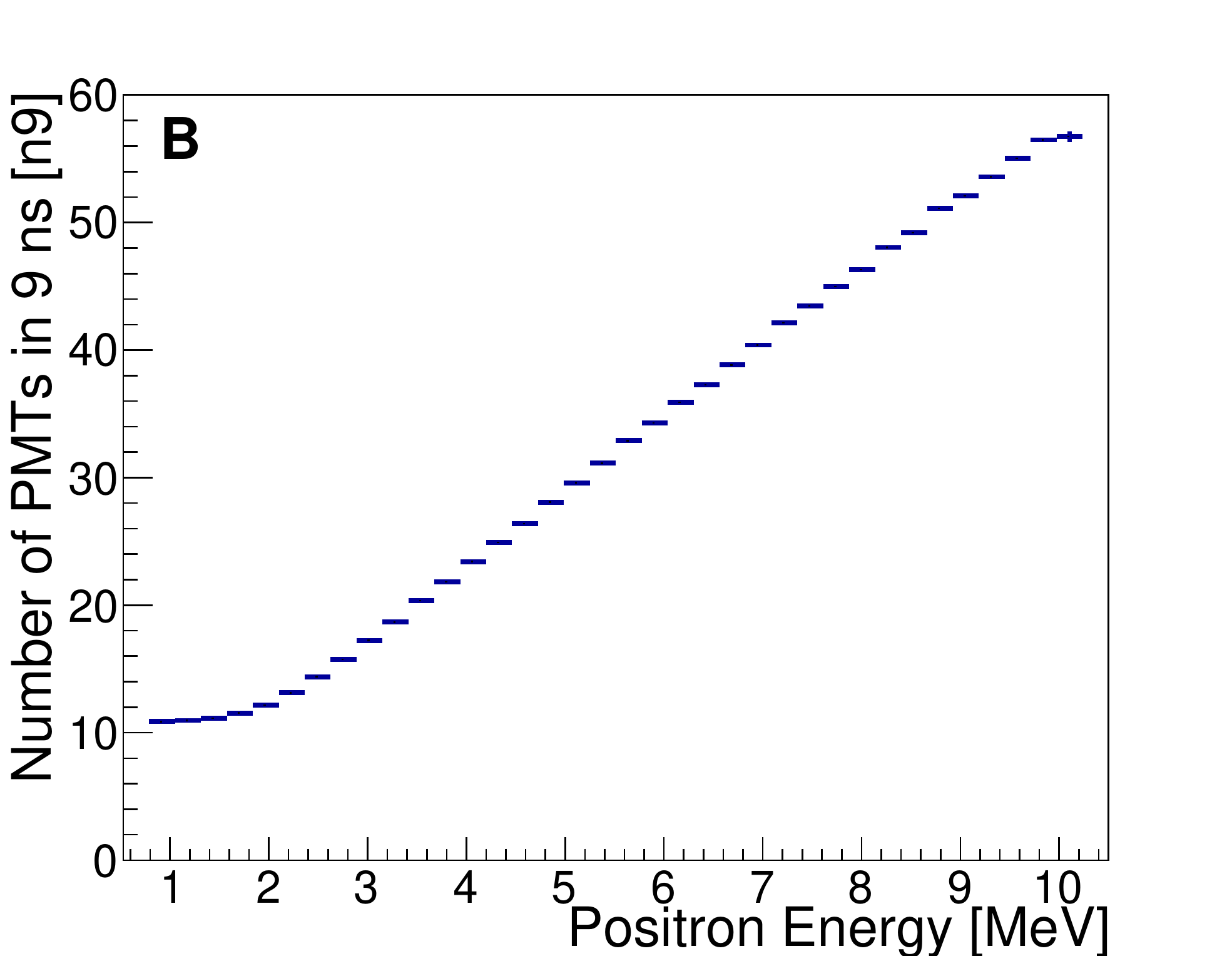}
\includegraphics[width=0.97\linewidth]{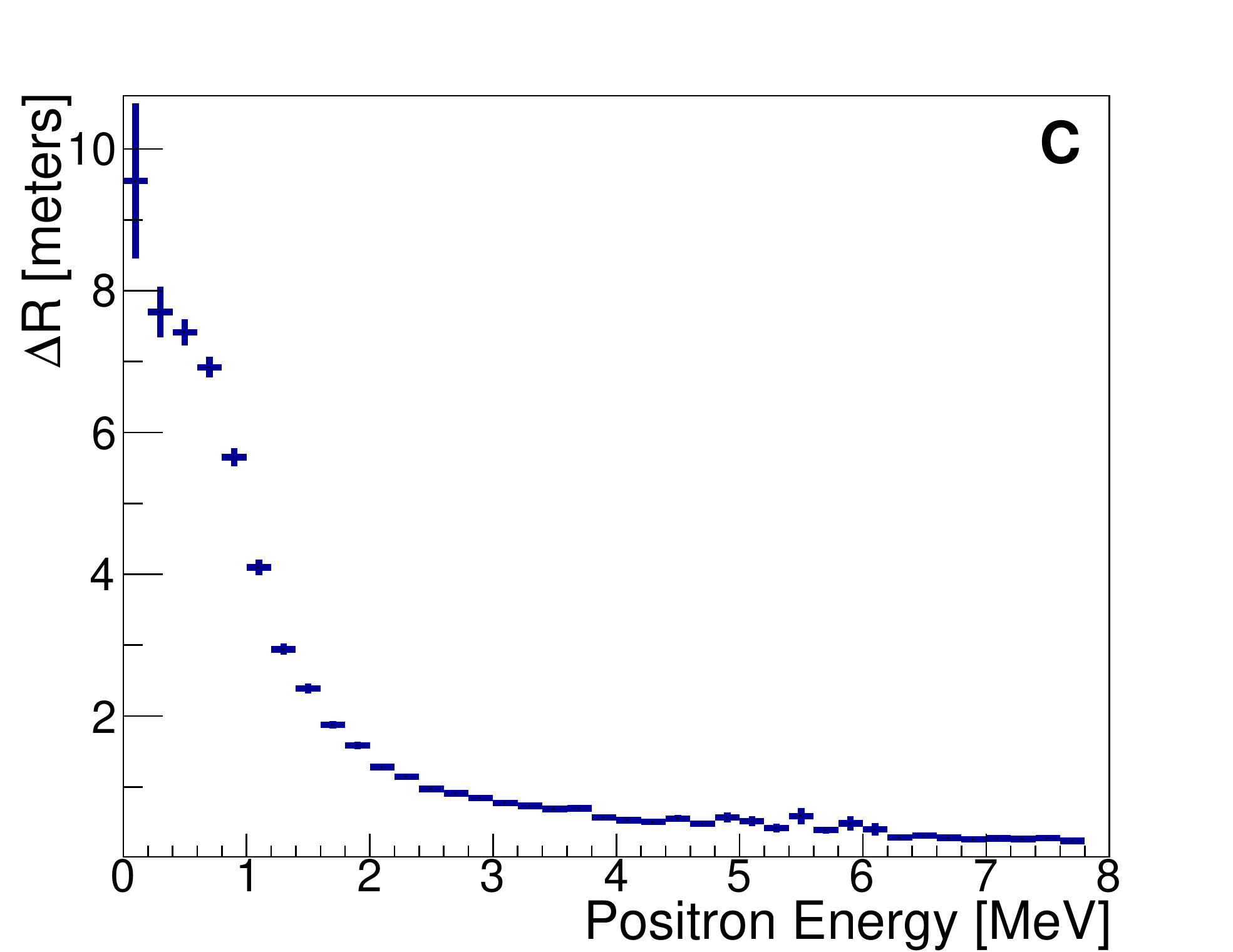}

\caption{ The PMT hit timing distribution from a set of 10\textsuperscript{7} events [A], the prompt PMT response to positrons as a function of energy [B], and the distance between true and reconstructed vertex as a function of energy [C]. 
}
\label{fig:positron_response}.

\end{figure}

\section{\label{sec:simulation}Simulation and Reconstruction Framework}
Various tools are used in the simulation, reconstruction, and event categorization to predict detector response. 
The Reactor Analysis Tool-Plus Additional Codes (RAT-PAC)\cite{ratpac}, employs Geant4 version 10.4 \cite{Geant4}, ROOT \cite{brun1997root}, and C++ to perform high fidelity Monte-Carlo simulations of the events expected from radiological processes, cosmogenically produced particles, and IBD. 
Following event production in the detector, all the subsequent processes are modeled, from optical photon production and transport through the water, to photon collection in PMTs.

The accuracy of the complex decay scheme resulting from neutron capture on Gd, as modeled in Geant4, has been an area of concern in recent years
\cite{yano2017measurement,hagiwara2019gamma}. 
Neutron capture on \textsuperscript{157}Gd generates the most prominent gamma ray cascade due to the larger cross-section relative to the other isotopes. 
To remedy the modeling concerns of Geant4, the DANCE \cite{DANCE}(Detector for Advanced Neutron Capture Experiments)Collaboration's gamma ray production results were incorporated into RAT-PAC using DICEBOX \cite{DiceBox}. 
The resulting simulation was validated against the WATCHBOY detector's neutron response and demonstrated better agreement with experimental data than the previous \textsuperscript{157}Gd gamma ray cascade model \cite{FastNeutron}. 

Following the simulation, the position and energy of each event are reconstructed using Branching Optimization Navigating Successive Annealing Iterations (BONSAI)
\cite{smy2007low}.
BONSAI uses the timing and the positions of the PMT hits to reconstruct the vertex position of each event based on the detected Cherenkov light and has been used by the Super Kamiokande collaboration for low energy event reconstruction \cite{sk_bon}. 

Antineutrino interactions are characterized by simulating the response to positron and neutron pairs. 
A flat positron energy spectrum was simulated to understand the energy-dependent detector response.
The emitted Cherenkov light from positron interactions results in approximately 9 detected photo-electrons (PE) per MeV. To inprove the energy resolution without increasing contributions from backgrounds, the energy proxy is taken as the number of PMTs that trigger from the prompt Cherenkov light in the first nine nanoseconds.

The energy-dependent positron response is seen in Figure \ref{fig:positron_response}. 
Figure 2.A shows that the PMT hit timing distribution from positrons
primarily occurs within a 9 ns window following  a start time defined by the event vertex. 
The negative times observed and the full width of the peak are due to the 2 ns PMT timing jitter, while the late light is caused by scattering in the detector. 
The features around 40-70 ns are caused by after-pulsing in the PMTs.
Figure 2.B shows that positrons below $~$1 MeV do not produce events above the rate of events seen from the dark rate in the PMTs. 
Lastly, once the prompt PMT hits are registered, BONSAI is used to reconstruct the events. 
Below 2 MeV, the  the true vertex is reconstructed to 1 meter; above 3 MeV the vertex reconstruction improves and rapidly converges at 50 cm, as shown in Figure \ref{fig:positron_response}.C.

\section{\label{sec:signal}Antineutrino Events of Interest}
After event reconstruction the individual events were sampled based on the reactor thermal power and standoff from the detector, $L$. 
The expected detected number of antineutrinos $N(E_{\overline{\nu}_e})$ is given by:

\begin{equation}
    N(E_{\overline{\nu}_e},L)=\frac{n_p T}{4\pi L^{2}} 
    \sum_lN^{f}_l\phi_l(E_{\bar{\nu}_e})\sigma(E_{\overline{\nu}_e})P_{ee}(E_{\overline{\nu}_e},L).\label{eq:detection}
\end{equation}
Here, $n_p$ refers to the number of quasi-free protons and $T$ is the counting time of the experiment. 
The electron antineutrino survival probability due to oscillations and the IBD cross section are given by $P_{ee}$ and $\sigma$, respectively \cite{vogel1999angular,gonzalez2014updated}. 
The individual contributions of fissile isotopes $l$ is represented by the fissile fraction for the specific isotope $N^{f}_l$ and the unique spectra for that isotope $\phi_l$, which is assumed to follow the approximations taken from the Huber spectra determination \cite{huber2011determination}. 

In effect, neglecting contributions from the backgrounds, the measured antineutrino signal will be governed by the standoff distance, burn-up, and power of the facility being monitored. 
The power of the reactor can be addressed through simple scaling as the thermal reactor power is directly related to the sum of the energy of the individual isotopes undergoing fission: $^{235}$U (201.912 MeV/fission), $^{238}$U (204.997 MeV/fission),$^{239}$Pu (210.927 MeV/fission), and $^{241}$Pu (213.416 MeV/fission) \cite{duderstadt1976nuclear}.
The Hartlepool reactors are AGRs, with more frequent refueling outages than pressurized water reactors. 
Given that the time scale of the reactor cycles for these reactors is on the order of weeks, and the dwell time for this analysis is on the order of months, the fuel burn-up effect is neglected. 
The analysis assumes static fissile contributions from $^{235}$U, $^{238}$U, $^{239}$Pu, and $^{241}$Pu based on the average fissile fraction of the Hartlepool Reactor Complex.

The systematic uncertainties for the antineutrino events were assumed to persist from a combination of the oscillation parameters, the accuracy to which the reactor power can be determined, uncertainties in the IBD cross-section, and the reactor-antineutrino anomaly.  
\section{\label{sec:backgrounds}Backgrounds}
\begin{figure}[t]
\includegraphics[width=0.98\linewidth]{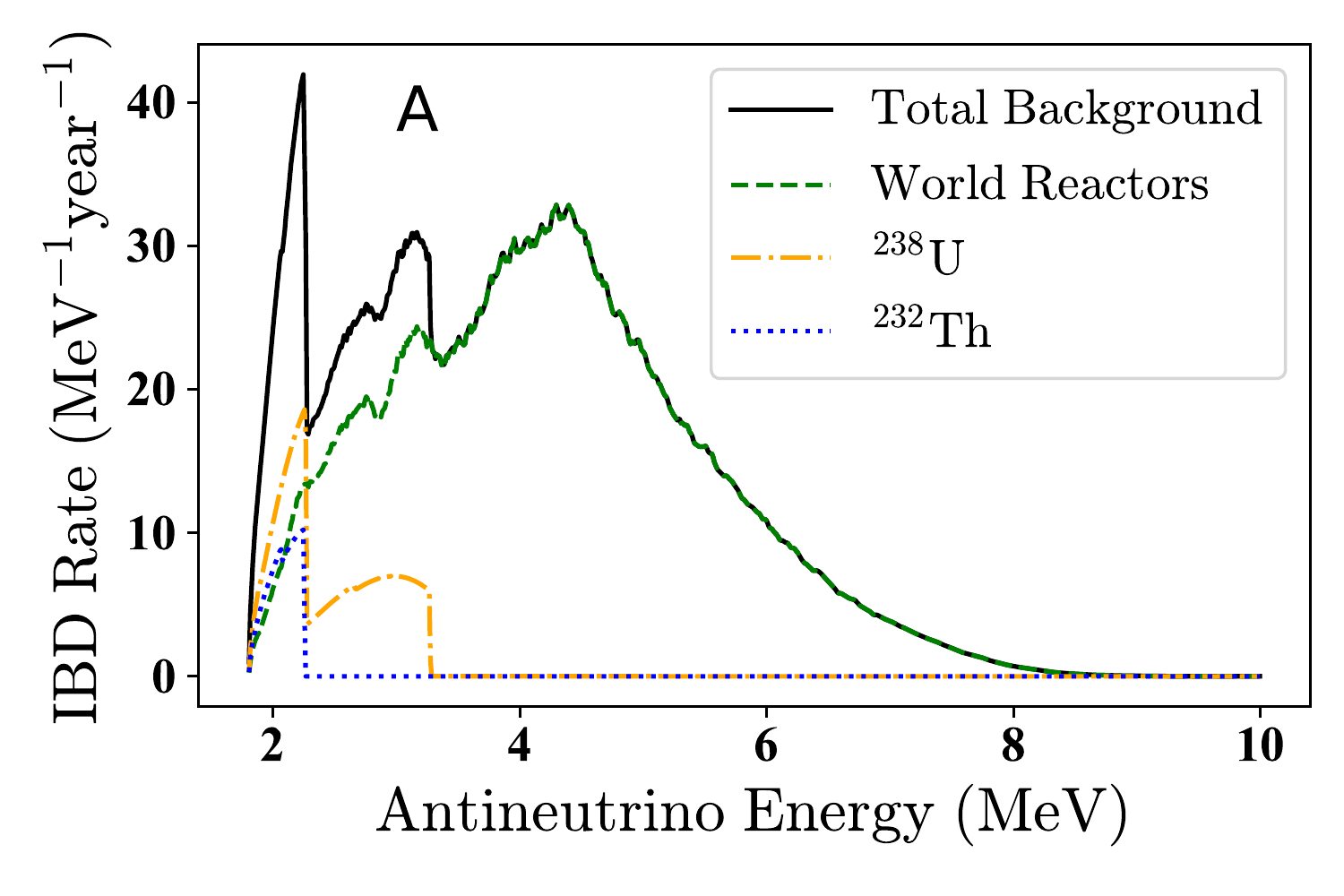}
\includegraphics[width=0.98\linewidth]{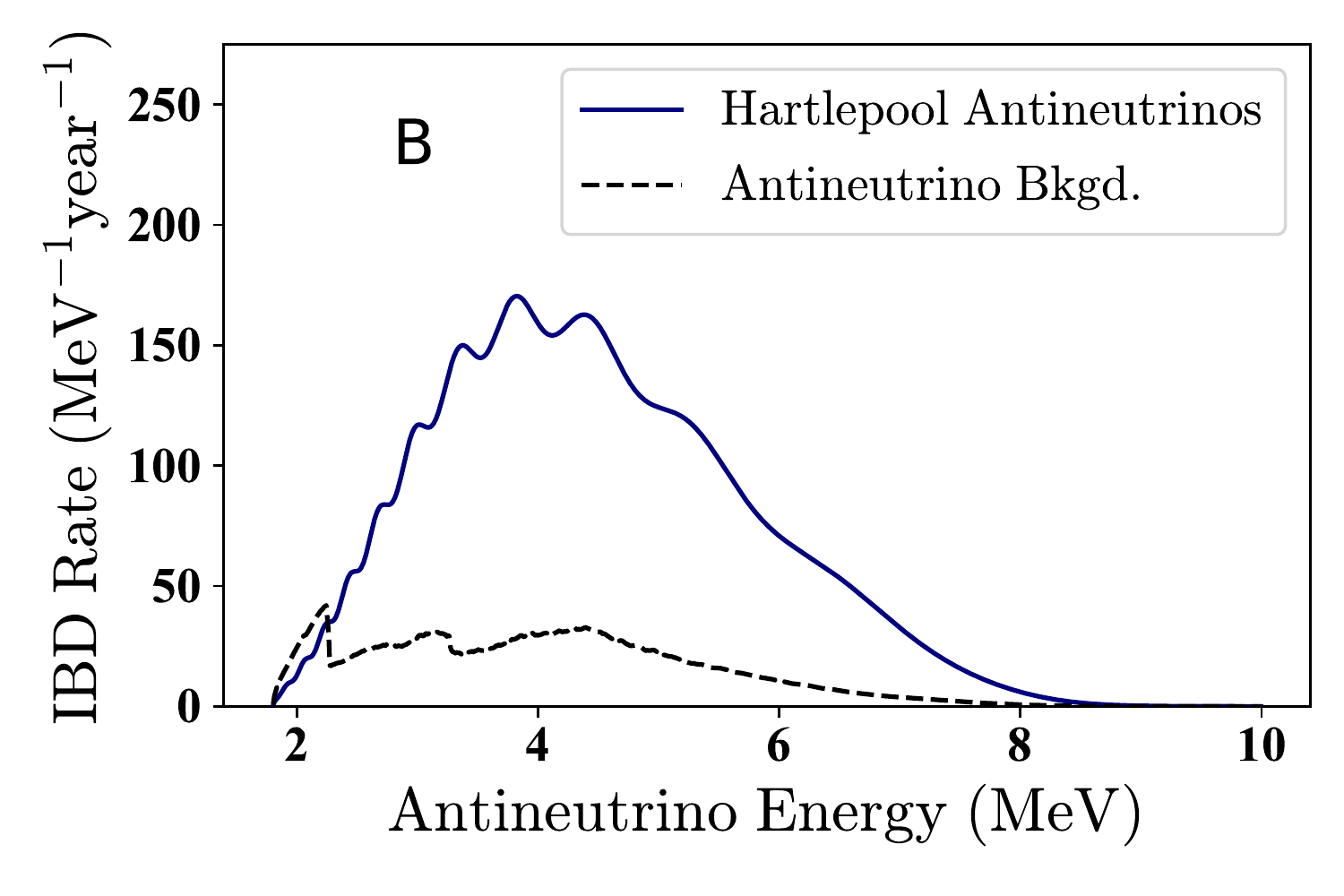}

\caption{The antineutrino backgrounds are shown in [A]. Here, the total expected number of antineutrino backgrounds (solid-black) from contributions from the world reactor backgrounds (dashed-green), and uranium (dashed-dotted yellow) and thorium (dotted blue) geoneutrinos. The reactor antineutrino flux from the Hartlepool Reactor Complex is shown in [B] along with  the antineutrino background.}
\label{fig:nu_back}
\end{figure}
Near-field antineutrino detectors dedicated to either monitoring reactor operation or 
neutrino oscillation physics have the benefit of using reactor ``off'' data to characterize and subtract their backgrounds \cite{PROSPECT}. 
Given the reactor discovery and timeliness goals for a far-field deployment, high-fidelity simulations are needed to understand when the recorded number of events is in excess of the modeled number of background events to a statistically significant degree. 
To do so, we divide our backgrounds into two categories: accidentals and correlated events. 
For the latter we consider contributions from other sources of antineutrinos, fast neutrons, radionuclide events, and the spontaneous fission of trace amounts of $^{238}$U and $^{232}$Th in the water, while the muon interactions in the detector are used to scale some of these correlated events and infer a deadtime. 
A detailed simulation of the detector components and surrounding materials is used to predict the rate of uncorrelated backgrounds.

\subsection{\label{sec:bkgd_wr}World Reactor Antineutrinos and Geoneutrinos}
Both geoneutrinos and the total world reactor antineutrino emission will contribute to the detected antineutrino background. Geoneutrinos are mostly emitted from the beta decay of uranium and thorium in the Earth's crust and mantle. Although emissions from $^{40}$K occur, their associated antineutrino energies are below the IBD threshold and thus, not detected. 
Additionally, the United Kingdom is bathed in a relatively high reactor antineutrino background compared to the rest of the world, with 15 installed nuclear reactors contributing to approximately 25 GW\textsubscript{th} of power. 
The total contribution of antineutrinos from the Hartlepool Reactor Complex in the Boulby mine will be 85$\%$. 
The next nearest reactor is the Heysham Reactor Complex at 149 km and will contribute 4.4$\%$ 
(a facility with approximately 6 GW\textsubscript{th} installed capacity), approximately 2.14$\%$ will come from geoneutrinos, while most of the remaining 8.5$\%$ of the antineutrino flux will come from other reactors in the UK and the large installed nuclear capacity in France. 

To quantify the antineutrino backgrounds, we take the non-Hartlepool antineutrino spectrum from geoneutrinos.org, a web-based application that uses the operating reactor capacity inputs from the Power Reactor Infotmation System (PRIS) database and couples the output with the Huber antineutrino spectrum, and a full treatment of oscillations \cite{barna2015global,calori1984iaea,huber2011determination}. The plotted contributions of these backgrounds are displayed in Figure \ref{fig:nu_back}. 

To account for the refueling outages of the various reactors, a 10$\%$ systematic uncertainty was applied consistent with the typical time in which reactors are not online. 
Additionally, the same considerations for the antineutrino signal of interest is used for the systematic uncertainty in the antineutrino background. Lastly, we incorporate the associated uncertainties described in Ref \cite{barna2015global}.

\subsection{Uncorrelated Backgrounds}\label{sec:accident}

\begin{figure*}[t]
\includegraphics[width=0.48\linewidth]{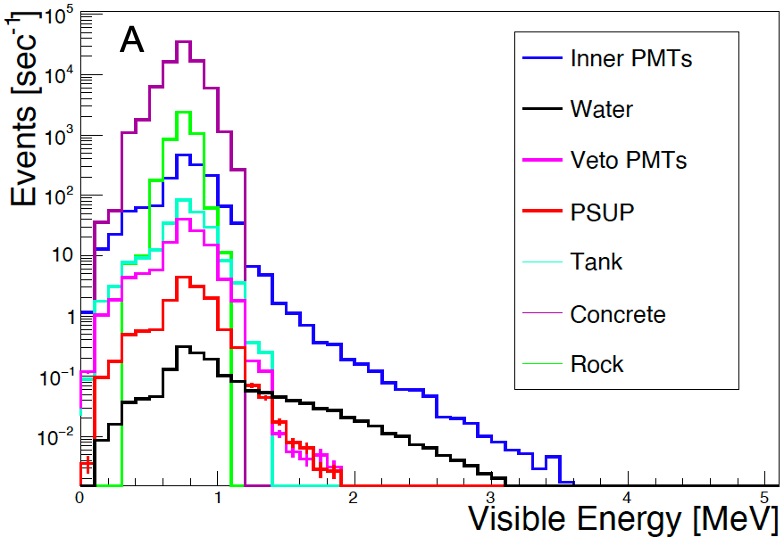}\includegraphics[width=0.48\linewidth]{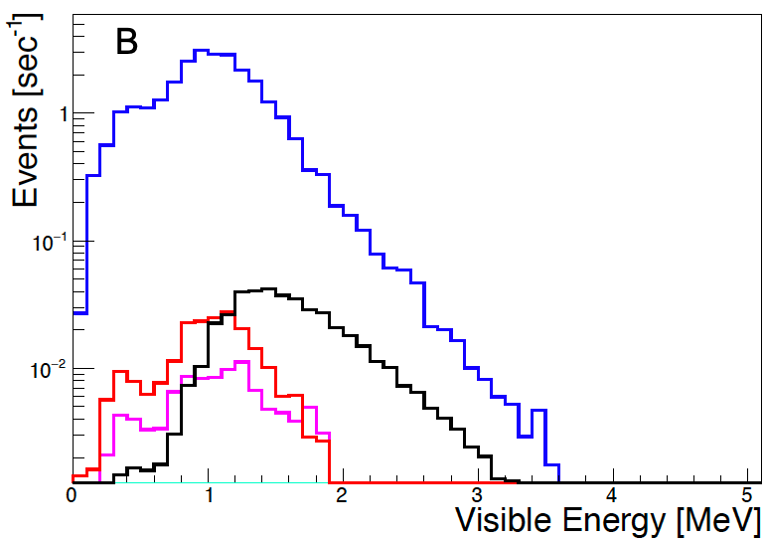}

\caption{[Color Online] Expected energy spectrum from the uncorrelated backgrounds in the detector without the trigger requirement [A] and after the PMT trigger requirement [B].
Contributions from the PMTs, water, PMT support structure (PSUP), tank, concrete, and rock are included. Despite large contributions of the concrete and rock layers around the cavern to the event rate, these events can be removed with a threshold requirement of 9 inner PMT hits within the trigger window of approximately 400 ns. After implementing PMT hit threshold, the largest contribution to the accidental background is due to the radiation from the PMTs. The accidental rate of the PMTs, PSUP, and the veto PMTs are reduced by constraining the fiducial volume within the tank to greater than 1.5-meter  from the PMT wall.  }
\label{fig:accident}.
\end{figure*}

Radiation from the decay of isotopes in the detector medium, PMTs, supporting structures, and surrounding materials will be emitted at random in the detector, providing a continuous uncorrelated background.
The event rate due to these backgrounds is expected to be high due to the large number of PMTs and other detector structures. There is a finite probability that accidental correlations will occur in close time and space coincidence, with similar coincidence signatures as IBD candidates. 
To quantify these backgrounds,
contributions from 
 radiological processes that can produce Cherenkov light were included in the simulation. 
The main decays included were: $^{232}$Th chain, $^{238}$U chain, $^{222}$Rn chain as part of the $^{238}$U chain, $^{40}$K, $^{60}$Co, and $^{137}$Cs. 
The 2019 WATCHMAN conceptual design reports details the process for which the radio-purity values of all components are determined and assumed \cite{bernstein2019ait}.
From here we assign a conservative 10$\%$ systematic uncertainty to our ability to precisely asses the radioactivity of all of the components. 
In a real experiment we can significantly reduce these values by measuring the singles rate and extrapolating to an assumed accidental background.
For the  $^{238}$U chain the following nuclides were simulated: $^{234}$Pa, $^{226}$Ra, $^{214}$Pb, and $^{214}$Bi. For the  $^{232}$Th chain the following nuclides were simulated: $^{228}$Ac, $^{224}$Ra, $^{212}$Pb, $^{212}$Bi, and $^{208}$Tl. 
In all cases it is assumed that the daughters in this chain are in secular equilibrium with their parent isotopes.

Radioactivity is simulated in both the inner and veto PMTs, the detection medium water, the PMT support structure, the detector tank, the concrete cavern, and the surrounding rock. 
Radiation production in the rock is only simulated to a thickness of 14 m, as it is expected that self-shielding dominates beyond that distance \cite{self_shielding}. 
The rate of accidental backgrounds in the detector, $R_\mathrm{acc}$, is expressed as:
\begin{equation}\label{eq:acc0}
    R_\mathrm{acc}=R_\mathrm{p}R_\mathrm{d}\Delta te^{-\Delta t(R_d+R_p)}\varepsilon_d.
\end{equation}
Here, $R_p$ is the expected prompt accidental signals and $R_d$ is the expected delayed accidental signal. The time and spatial coincidence required to satisfy an IBD-like event is set as $\Delta t$=100 $\mu s$ and $\varepsilon_d$=0.05, respectively. 
The spatial coincidence, $\varepsilon_d$, is the probability of 2 uncorrelated single events reconstructing within the 2-meter distance cut.
Simulations of the uncorrelated backgrounds were performed to evaluate the spatial coincidence probability which led to the assignment of a 20$\%$ systematic uncertainty. 
Given that $\Delta t(R_d+R_p)<<1$, the term in the exponential can be ignored, and Equation (\ref{eq:acc0}) becomes:
\begin{equation}\label{eq:acc1}
    R_\mathrm{acc}=R_{p}R_{d}\Delta t\varepsilon_d.
\end{equation}
Additionally, the contribution from alpha decay is not included in the simulation, as these particles are not expected to be relativistic in our detector and don't produce Cherenkov light. 
The simulated energy distribution of detector components with and without the analysis selection threshold are shown in Figure \ref{fig:accident}.

The largest impact to the uncorrelated rate is the radiation from the concrete and rock.
The contribution of the uncorrelated backgrounds to the veto detector are discussed in the next section. 
Most of these events occur in the outer veto region and produce very little light in the inner detector, hence they can be removed with a threshold requirement of 9 prompt inner PMT hits. 
After these analysis selction thresholds the strongest contributor to the uncorrelated event rate is the radiation from the PMTs, with lower order contributions from the veto PMTs, the water in the detector, and the PMT support structure (PSUP). 
Prior knowledge of this contribution motivated the use of low-radioactivity glass for the PMT selection. 
The contribution of PMT dark noise is simulated assuming a conservative rate of 10 kHz following Poisson statistics. 
The PMTs have a mean rate of 2 kHz; however, previous experiments with submerged PMTs has shown that the singles rate can increase when placed in water or scintillator \cite{PMTdata}.

\subsection{\label{sec:muons}Muons}
\begin{figure}
    \includegraphics[width=0.98\linewidth]{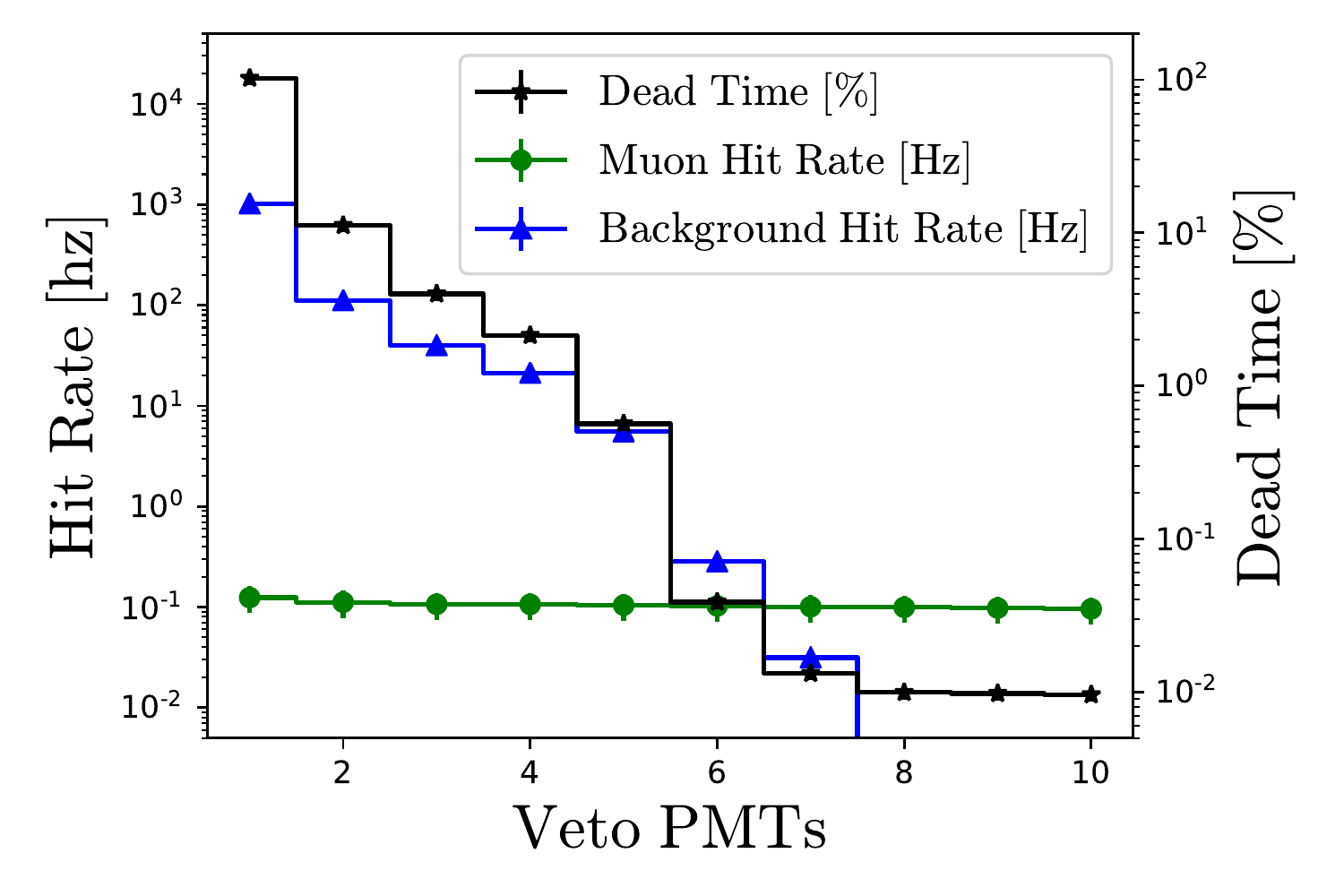}
    \caption{Rate of triggers in the veto detector due to muon interactions [green dots], radiation in the detector and surrounding material [blue triangle], and the resulting dead-time  [black star]. Up to a 10 veto PMT hit threshold, the rate of triggers from muons is relatively constant, while the rate of triggers from radiological backgrounds becomes negligible at a threshold of 8-PMT hits.}
    \label{fig:vetorate}
\end{figure}

Aside from the antineutrino background, the highest contribution to correlated backgrounds originates from the progeny of cosmic-ray muons in the detector and surrounding materials. 
The veto region is used to identify and reduce the contribution of such backgrounds. 
The veto PMTs are facing outwards relative to the inner detector, and have a photo-coverage of approximately 1.5$\%$. 
The muogenic response of the detector is determined by simulating the energy and directional distribution from a parameterization with a normalization to the overall muon flux, $I_{\mu}$ in units of cm$^{-2}$s$^{-1}$ \cite{mei2006muon}: 
\begin{eqnarray}
    I_{\mu}(h)=(67.97e^{-h/0.285} +2.071e^{-h/0.698})\times10^{-6} 
    \label{eq:mh_muflux}
\end{eqnarray}
The muon flux is determined by the depth in km.w.e., $h$, and shape of the overburden.
The overburden has a depth of 2.805$\pm$0.015 km.w.e. (kilometer water equivalent).  
Using Equation (\ref{eq:mh_muflux}), the expected muon flux is expected to be (4.09$\pm$0.15)$\times10^{-8}$ cm$^2$ sec$^{-1}$. The integrated muon flux was simulated as 0.116$\pm$0.004 sec$^{-1}$.
In addition to the rate of muons, the response in the detector is highly correlated to the muon energy and direction.
The energy distribution of the muon flux, $ \dfrac{dN}{dE_{\mu}}$, is assumed to be:
\begin{equation}
    \dfrac{dN}{dE_{\mu}}= Ae^{-bh(\gamma_{\mu}-1)}(E_{\mu}+\epsilon_{\mu}(1-e^{-bh}))^{-\gamma_{\mu}}.
    \label{eq:me}
\end{equation}
The angular distribution of the muons is taken from Ref \cite{mei2006muon}, and is described by the angular  intensity,  $I(\theta)$:
\begin{equation}
    I(\theta,h)= \sec\theta \times e^{-h/\gamma(sec\theta-1)}.
    \label{eq:mt}
\end{equation}
Here, $\gamma_{\mu}$=3.77, $\epsilon_{\mu}$=693 GeV, and $b$=0.4/km.w.e., are described in detail in Ref. \cite{mei2006muon}. 
The values are the result of fitting multiple experiments to a generalized form. 
Previous experiments have demonstrated reasonable agreement between the simulated and observed muon response \cite{sno_muons}.
After the muons were simulated, we determined the veto criteria by evaluating the response to muon events in the detector as well as the response from uncorrelated backgrounds that may cause the veto to trigger. 
For this study we want to maximize the veto trigger rate from muons, while minimizing deadtime due to backgrounds. 
Figure \ref{fig:vetorate} shows the trigger rate in the veto detector as a function of the threshold PMT hits required to define a muon event.
Following a muon trigger, the experience gained from the WATCHBOY detector suggests a 1 ms deadtime is needed to remove the contributions from Michel electrons, after-pulsing, and muon induced neutrons \cite{dazeley2016search}. 
If we assume an 8-veto-PMT-hit threshold, and a 1 ms deadtime following a trigger the total deadtime of the detector is expected to be $0.98 \pm 0.04\%$. The detection efficiency for muons in the veto detector is taken as 100$\%$.
The contribution of undetected muons to the background will be due to the induced rate of  di-neutrons or the radionuclides generated within 1 ms of their paramuon.

\subsection{\label{sec:fastneutrons} Muogenic Fast Neutrons}

Although the veto region enables tagging of muon-induced fast neutron contributions in the detector, its ability to reject fast neutrons  produced in the rock is limited. The veto region does act as a buffer region to protect the inner detector from these neutrons.
Although fast neutrons generated in the rock can reach a multiplicity of up to 20, these events can mimic an IBD event if two neutron captures are detected within the prescribed time and spatial separation cuts. 
Higher order coincidences, or events with 3 or greater coincidences, will be rejected; however, the di-neutron contribution is an irreducible background and must be quantified to a high fidelity.

Fast neutrons' inelastic collision with nuclei are poorly modeled by certain simulation frameworks. A concern arose that the production of high neutron multiplicity events in Geant4 may be inconsistent with measured data.
Indications that our Geant4-based simulation (version 4.9) may be inconsistent with measured data were also found in comparisons with  experimental data from the WATCHBOY detector \cite{FastNeutron}. The WATCHBOY detector was deployed in KURF (Kimballton Underground Research Facility) for the purpose of validating the muogenic response of fast neutrons and radionuclides. 
Although the detector was not sensitive enough to measure the \textsuperscript{8}He and \textsuperscript{9}Li production, a detailed measurement of the fast-neutron multiplicities in the rock was generated and confirmed to be inconsistent with the Geant4 simulation's results.
Alternative simulations of the WATCHBOY detector were also performed using FLUKA (FLUktuierende KAskade). 
As outlined in Sutanto \textit{et al.}, the results of the FLUKA simulation agreed with experimental results and validated the use of this code making predictions for the Gd-H$_2$O detector design \cite{FastNeutron}. 

\begin{figure}
    \includegraphics[width=0.98\linewidth]{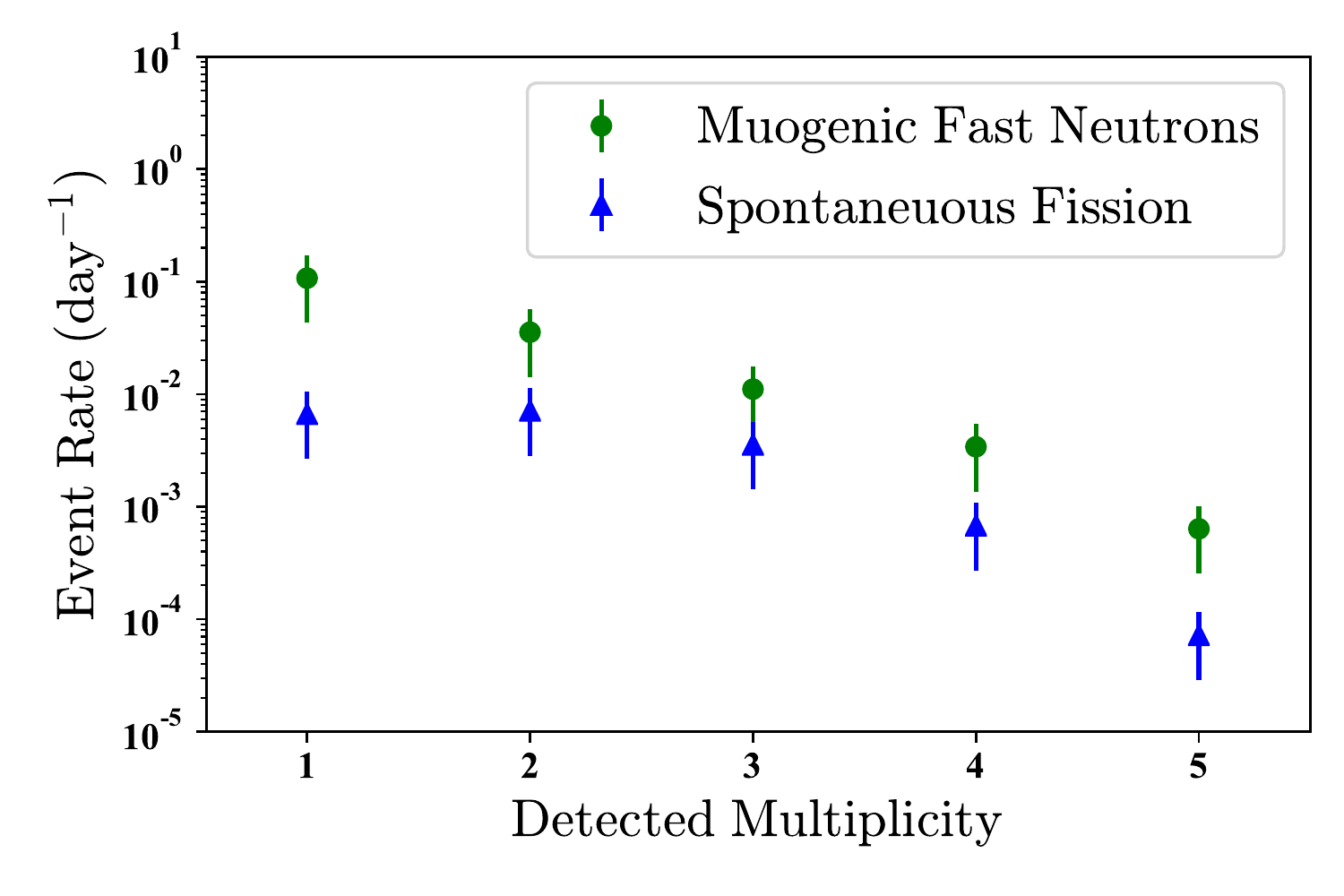}
    \caption{Cluster events as a function of multiplicity from the fast muogenic neutrons [green dots] and the spontaneous fission of $^{238}$U and $^{232}$Th in the detector [blue triangles]. The fast neutron events was modeled using FLUKA, while the spontaneous fission of $^{238}$U and $^{232}$Th was modeled using FREYA \cite{FastNeutron,FREYA}. }
    \label{fig:dineutron}
\end{figure}

The muon production described in the Mei and Hime study was simulated in FLUKA for tracking the fast neutron production and multiplication in the rock and detector \cite{battistoni2007fluka}. 
The detector design, cavern geometry, and rock composition were imported into the FLUKA simulation to match the geometry of the Geant4 simulation. 
To avoid the production of neutrons in Geant4 from simulated fast neutrons in FLUKA while allowing for a consistent detector response with the other simulated backgrounds, the histories of muogenic neutrons were traced until their capture location in the detector. 
Once the capture location was recorded from the FLUKA simulation, thermal neutrons at the same location were simulated in Geant4 to produce a realistic response to neutron capture and a consistent spatial reconstruction of events.
 From Figure \ref{fig:dineutron}, the detected di-neutron rate in the fiducial volume is expected to be 0.0356 events per day. A 40$\%$ systematic uncertainty on the di-neutron rate was assigned due to the difference between the WATCHBOY measurement and the FLUKA simulation \cite{FastNeutron}.

\subsection{\label{sec:radionuclides}Muogenic Radionuclides}
\begin{figure}[ht]
    \includegraphics[width=0.98\linewidth]{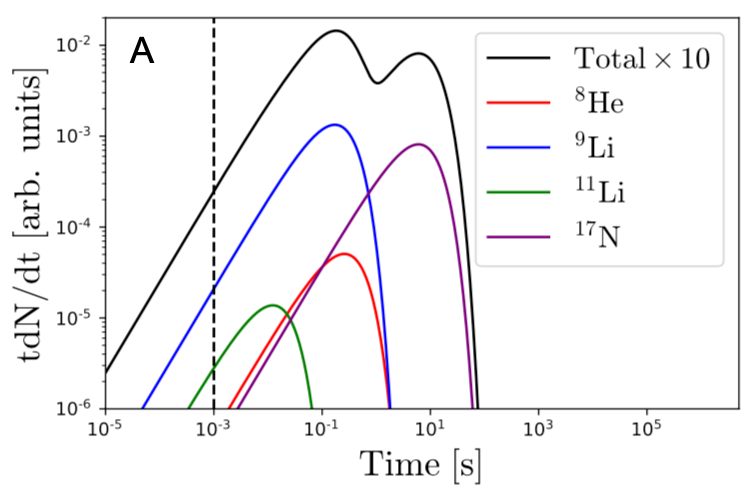}
    \includegraphics[width=0.98\linewidth]{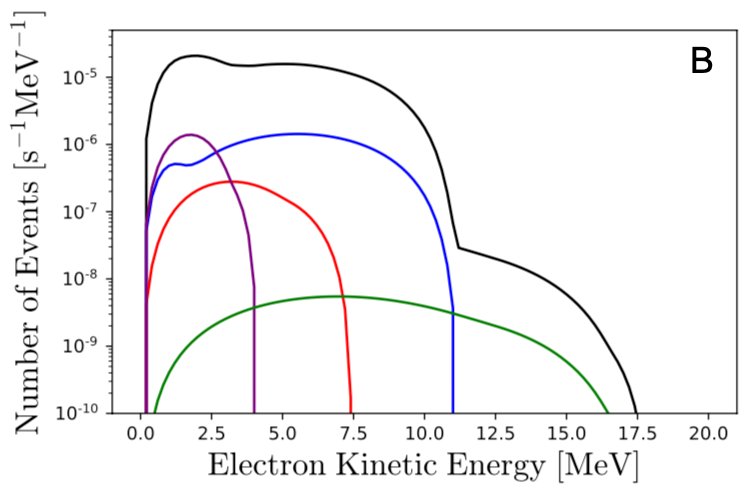}
    \caption{[Color Online] Decay timing [A] and energy distribution [B] of the radionuclides produced from  muon events. The decay time distribution is unitless and the height of the curves shows the relative distributions of event. It can be seen that a majority of the radionuclides are generated past the muon veto deadtime of 1 ms. The electron kinetic energy spectra was generated from literature values of the beta endpoints following decay and are shown in [B]. Given the 2 MeV energy threshold in this design, contributions from $^{17}$N and $^{8}$He are minimized.}
    \label{fig:radionuclides}
\end{figure}

Radionuclides present an irreducible background that occurs in the detector from muon spallation. The isotopes decay through the emission of a beta particle to the excited state of the daughter nucleus. 
For the nuclei that are excited beyond the neutron separation energy, the immediate evaporation of a correlated neutron can mimic an IBD event. These so called $\beta$n reactions can have lifetimes that exceed a reasonable veto-induced deadtime. 
The production of radionuclides as a function of muon path length has been calculated by Li and Beacom using FLUKA simulations \cite{li2014first}. 
The difference in the observed and predicted radionuclide production in Super Kamiokande was used to identify the systematic uncertainty in using this parameterization \cite{SuperK}. 
For any water-based detectors, $\beta$n reactions from the decay of $^8$He $(t_{1/2}= 0.178 \ \mathrm{ sec})$, $^9$Li $(t_{1/2}= 0.119 \ \mathrm{ sec})$, $^{11}$Li $(t_{1/2}= 0.0085 \  \mathrm{ sec})$, and $^{17}$N $(t_{1/2}= 4.173 \  \mathrm{ sec}) $ are of concern. Theoretically, contributions from $^{16}$C are possible; however, due to their low production rate they are assumed negligible for the present design.

The time dependence and beta-energy spectrum from these radionuclides are shown in Figure \ref{fig:radionuclides}.
Despite being correlated with muon events in the inner detector, these backgrounds persist past the 1-ms deadtime due to the the lifetimes of these isotopes.
A low-energy threshold would significantly reduce the contributions from $^{17}$N given the endpoint energy of the largest $\beta$n branches: 4.1 MeV (BR=37.5$\%$) and 3.3 MeV (BR=49.9$\%$). 
To understand the detector response to these backgrounds, electrons sampled from the beta spectra, and accompanying gamma rays were simulated to determine the prompt response. The neutron produced from this process was assumed to be emitted isotropically. The shape of the beta spectrum is described by:
\begin{equation} \label{eq_beta}
    N(T) = C(T) F(Z,T) p E (Q-T)^2,
\end{equation}
where $T$ is the kinetic energy, $C(T)$ is the shape function (in this study, only the allowed decay is considered and hence this factor is a constant), $F(Z,T)$ is the Fermi function used to correct for electron screening, $Z$ is the charge of the final nucleus, $p = \sqrt{(E/c)^2 - (mc)^2}$ is the momentum,  $E =  T + mc^2$ is the total beta energy, and $Q$ is the energy released for each decay.

In addition to the spallation rates, the lifetimes of the radionuclides were used to determine the rate of interactions.
Radionuclides that decay within 1 ms of the muon event will be rejected with close to 100$\%$ due to the imposed deadtime of the detector. 
Additionally, it is assumed that radionuclides can be tracked with a 90$\%$ efficiency, resulting in a 90$\%$ rejection for radionuclides that decay within 1 ms to 1 sec after muon interactions.
When combining timing cuts with the detector response we expect to observe 0.225 IBD candidates per week from $^{17}$N, 0.098$\pm$0.025 per week from $^9$Li, 0.003$\pm$0.002 per week from $^8$He, and 0.001$\pm$0.001 per week from $^{11}$Li.

\subsection{Spontaneous Fission\label{sec:SF}}
In addition to the contribution of accidental backgrounds in the detector, trace amounts of $^{238}$U and $^{232}$Th in the water can create correlated events through spontaneous fission. 
A spontaneous fission event will result in a prompt gamma ray flash and the emission of multiple neutrons. 
Spontaneous fission may mimic an IBD event if any one of the prompt gamma ray events, or a neutron capture
reconstructs as a positron candidate followed by the detection of one fission neutron.
As in the case of the fast neutron contamination of the background, events with detected multiplicities greater than two are neglected. 

To effectively evaluate the contribution from spontaneous fission, 
the momentum, energy, and multiplicities were calculated using  Fission Reaction Event Yield Algorithm (FREYA) \cite{FREYA}, an useful tool for calculating event-by-event correlations in fission observables. 
Figure \ref{fig:dineutron} shows the probability distribution of detected multiplicities for both $^{238}$U and $^{232}$Th for spontaneous fission in the detector. 
These values are plotted with the detected fast neutron multiplicities because these backgrounds can be rejected when additional events are recorded in that time window. 
In total, the rate of IBD candidates from spontaneous fission of \textsuperscript{232}Th and \textsuperscript{238}U is expected to be $2.60\pm 1.62$ events a year for an assumed uranium and thorium concentration of 7.0 Bq and 0.9 Bq, respectively, in the 6.3 kiloton purified Gd-doped water volume.

\section{\label{sec:summary}Event Summary}

\begin{table}[t]
    \protect\caption{Signal and background rates for the Gd-H$_2$O baseline detector design, assuming deployment at the Boulby site.  Rates include the signal from the Hartlepool Reactor Complex and all backgrounds evaluated in this study. The italicized values correspond to the statistical and systematic uncertainties. }
\begin{ruledtabular}
\begin{tabular}{lll}
Contribution & Events [year$^{-1}$] & $<\sigma>$/N \\
\hline 
Reactor Signal         & 354.32 $\pm$31.39& \\
\textit{Statistical} & & 5.3$\%$\\
\textit{Thermal Power} & & 5$\%$\\
\textit{IBD Cross Section} & & 0.5$\%$\\
\textit{Neutrino Flux} & & 5$\%$\\
\hline
Antineutrino Background   & 47.41 $\pm$9.01& \\
\textit{Statistical} & & 14.5$\%$\\
\textit{Capacity Factor} & & 10$\%$\\
\textit{geoneutrinos.org}\cite{barna2015global} & & 5$\%$\\
\textit{Neutrino Flux} & & 5$\%$\\
\hline
Accidentals           & 32.40 $\pm$9.21& \\
\textit{Statistical} & & 17.6$\%$\\
\textit{Radioassay} & & 10$\%$\\
\textit{Voxelation} & & 20$\%$\\
\hline
Muogenic Radionuclides          & 13.73 $\pm$7.80& \\
\textit{Statistical} & & 26.9$\%$\\
\textit{Model Validation}\cite{li2014first,zhang2016first} & & 50$\%$\\
\hline
Muogenic Dineutrons             & 13.00 $\pm$6.32& \\
\textit{Statistical} & & 27.7$\%$\\
\textit{Model Validation}\cite{FastNeutron} & & 40$\%$\\
\hline
Spontaneous Fission          &2.60 $\pm$1.62& \\
\textit{Statistical} & & 62$\%$\\
\textit{Model Validation} & & 3$\%$\\
\hline 
Total & 463.46 $\pm$65.35\\
\end{tabular}
\end{ruledtabular}
    \label{tab:event_summary}
\end{table}

\begin{figure*}[t]
\includegraphics[width=0.49\linewidth]{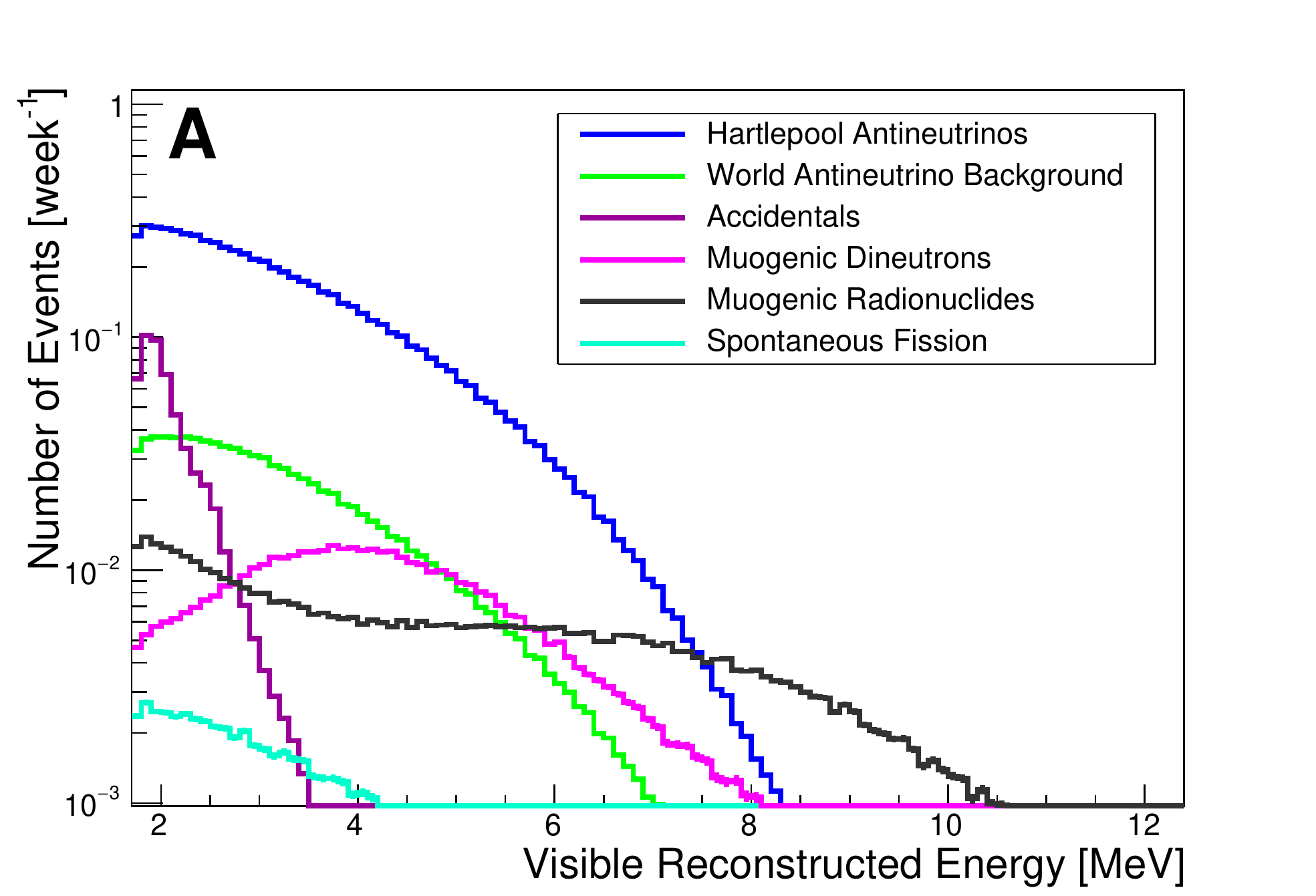}\includegraphics[width=0.49\linewidth]{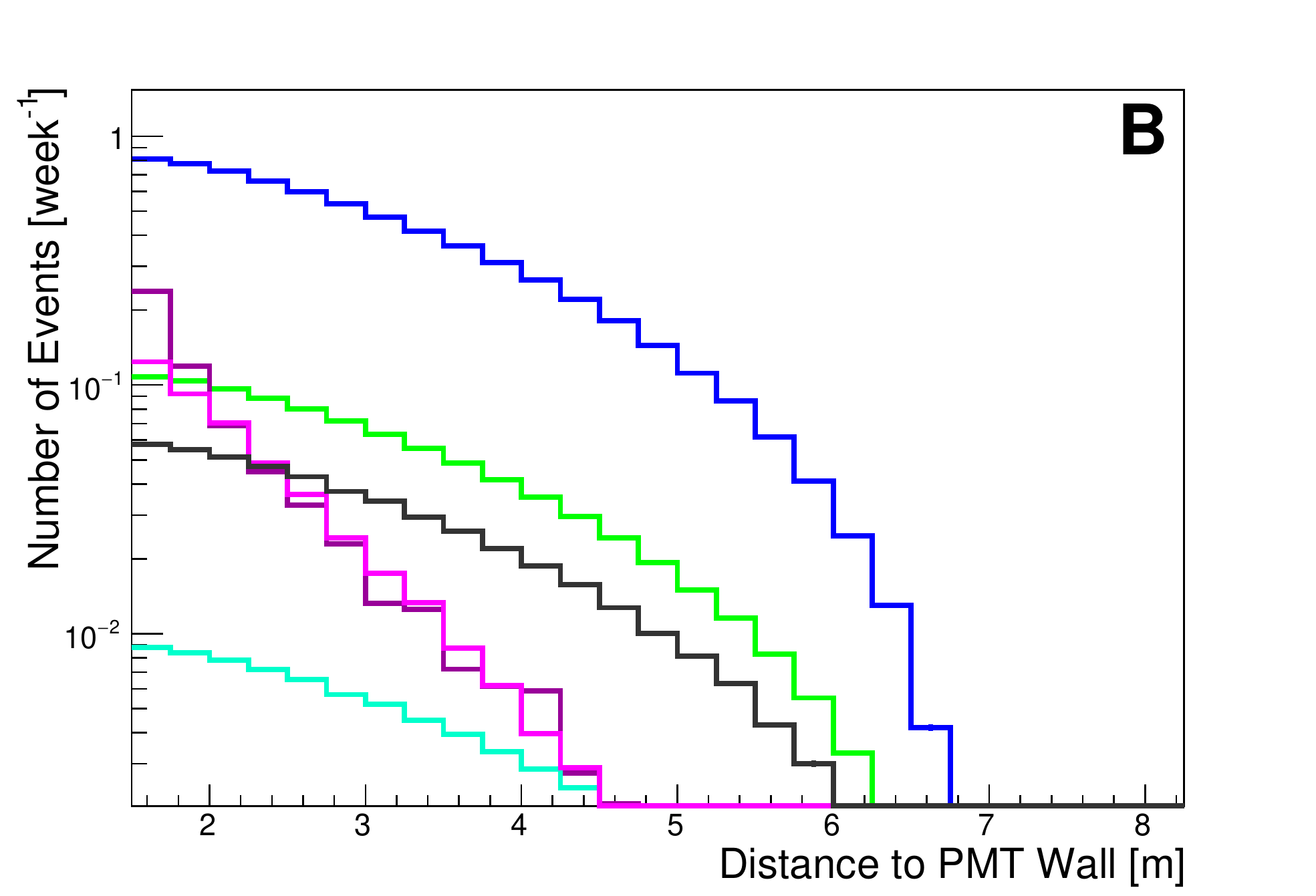}

\includegraphics[width=0.49\linewidth]{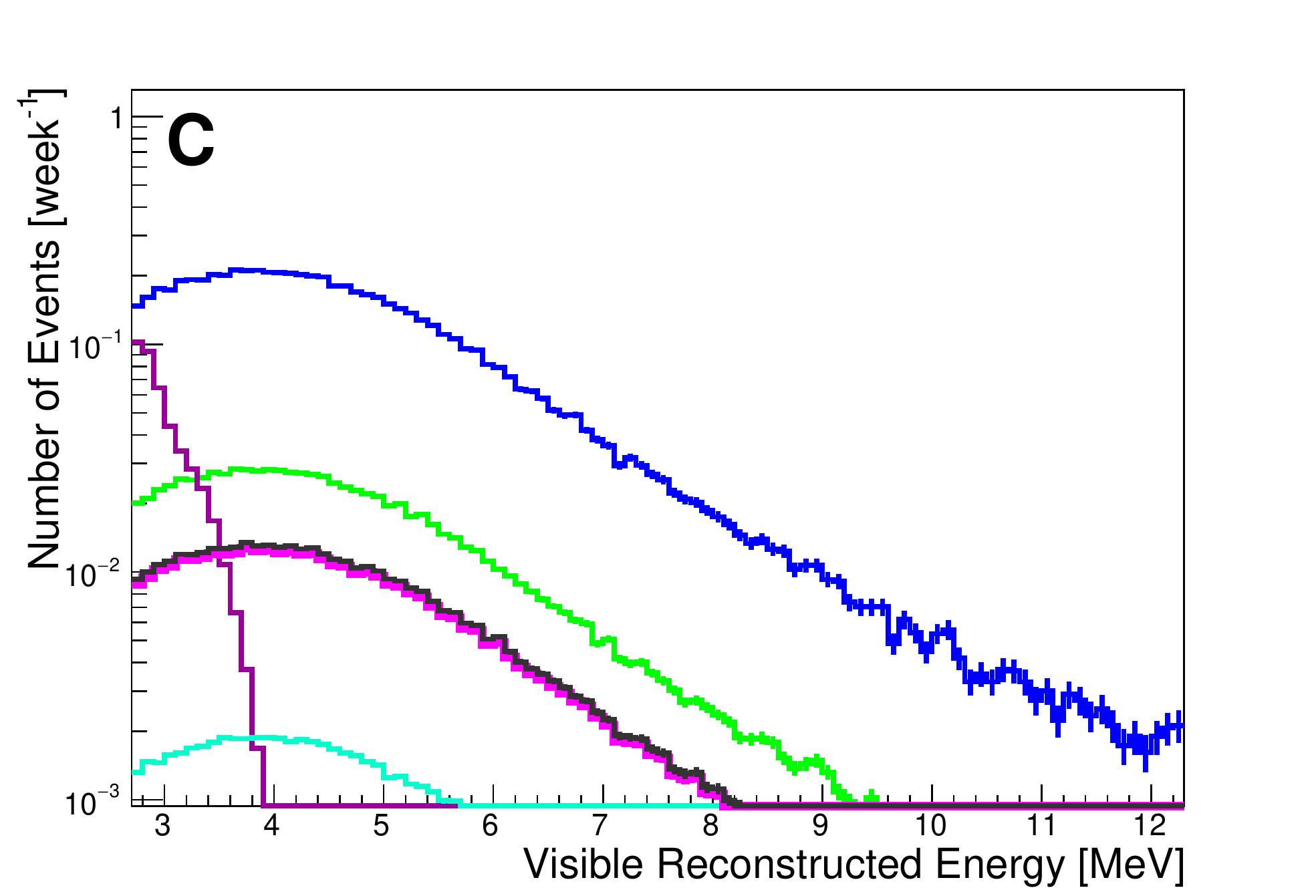}\includegraphics[width=0.49\linewidth]{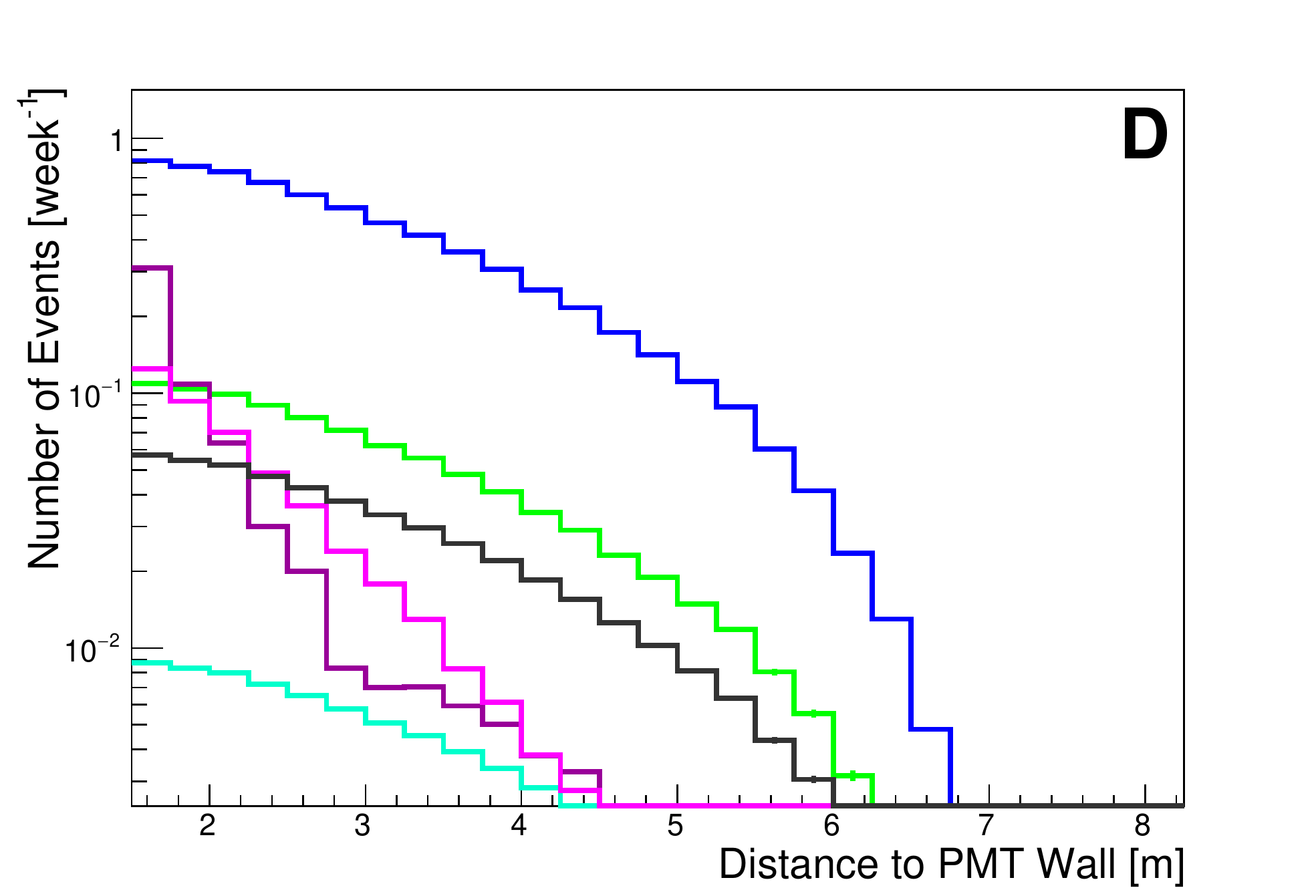}

\caption{[Color Online] The energy and spatial distribution of the prompt events [A and B ] and the delayed signal [C and D]. Here, the contributions from the Hartlepool Reactor Complex (26 km standoff distance and 3 GW\textsubscript{th} power), the world reactor background, the accidentals, muogenic dineutrons, muogenic radionuclides, and spontaneous fission are shown. A 2 MeV threshold is required for prompt events, while a 3 MeV threshold is required for delayed events. All events must reconstruct 1.5-meters away from the PMT wall. }
\label{fig:totevent}.
\end{figure*}
After setting a threshold of 2 MeV for the prompt positron, 3 MeV for the delayed neutron, and a 1-kton fiducial volume. The greatest contributor to the background is the world reactor antineutrino flux, followed by the coincident rate of accidentals. 
The next largest backgrounds are muogenic radionuclides and fast neutrons. 
A summary of the background rates is found in Table \ref{tab:event_summary}.  
Although this study evaluates variable reactor power and standoff distances, the signal rate shown is for the Hartlepool Reactor Complex at full reactor power (i.e a nominal 26 km standoff distance and 3 GW\textsubscript{th} reactor power).

Two parameters in the detector response that have a distinct feature for both the signal and background are the charge distribution profile, or in this case the visible energy; and the location where the event is reconstructed.
In the reactor exclusion case discussed in Section \ref{sec:nonpro}, the main goal is to identify when events can be seen above background. 
With this in mind, the two largest contributors to the background are the accidentals and the reactor antineutrino background. 
A large fraction of the accidentals is attenuated by the buffer region, limiting their charge deposition to the lower portion of the spectrum and constraining the event reconstruction to events closer to the PMT wall. 
Although antineutrinos from the world reactor background will have the same spatial response as the signal of interest, distortions in the spectral response may occur due to the oscillation of the signal. 
The same distortion from oscillations in the number of PMT hits is utilized to attempt to determine the reactor power and standoff distance in the verification case discussed in the next section. Figure \ref{fig:totevent} shows the spatial and energy distributions of the signal and background events.

The values represented in Table \ref{tab:event_summary} include both statistical and the systematic uncertainties. 
In the absence of a physical detector, broad assumptions were made when evaluating the systematic uncertainties based on model inputs, previous measurements, and comparison studies \cite{FastNeutron,dazeley2016search,roecker2016design,li2014first}. 
However, since the signal event rate in this design is high compared to most backgrounds (apart from other reactors and accidentals) the systematic uncertainties on those other backgrounds are negligible compared to the statistical and systematic uncertainties for the two main backgrounds.

\section{\label{sec:nonpro}Sensitivity Methodology}
This study evaluates the sensitivity of the Gd-H$_2$O baseline detector design for two specific cases:
\begin{enumerate}
  \item Excluding the presence of undeclared nuclear reactors from a given region.
  \item Confirming the reactor power and standoff of a declared nuclear facility.
\end{enumerate}
In Case 1 the Hartlepool Reactor Complex is assumed to not exist, but associated backgrounds calculated at the Boulby underground laboratory are still used. 
For Case 2, the reactor search and ranging capability of the  detector is calculated for the Hartlepool complex. 
In both cases, the study examines the detector's ability to exclude various reactor powers and standoffs based on a set of pre-determined conditions.

The sensitivity is evaluated by first determining the two-dimensional PDF for the detector response as a function of energy deposition and spatial reconstruction vertex for both the prompt and delayed signals. 
The spatial reconstruction of the prompt positron and neutron are correlated and sampled as such. 
From there, simplified Monte Carlos are used to identify the confidence limits by fitting against the PDF scaled with the experiment time. 
As different standoff distances and reactor powers are probed, simplified Monte Carlos are fit against the hypothesis template. 
The results of which determine the confidence level for Case 1 and 2.

Following closely the analysis procedure detailed in Albert \textit{et al.}, a template, $T$, for the no-reactor case and the Hartlepool reactor experiment is generated by sampling the two-dimensional PDF of the signal and background event reconstruction for the energy response, $E$, and the reconstructed event location, $\bar{r}$, for both the prompt signal, denoted by $p$,  and delayed signal, denoted by $d$, IBD candidate pair for the signal and background using Equation (\ref{eq:PDF}) \cite{NEXO}:
\begin{equation}\label{eq:PDF}
    T^{p,d}(\bar{r},E)=\sum_i N_iR^{p,d}_i(\bar{r},E)\times t.
\end{equation}
Here, $i$ refers to each process that can generate IBD candidates, $N$ refers to the event rate of each process, $R$ refers to the reconstructed response to each event, and $t$ is the experiment time.  
These templates are used to understand  sensitivity  to deviations from the assumed reactor operations. 

 For each reactor power, detector standoff, and counting time; a simplified data set was generated by randomly sampling the PDF of event reconstruction for the charge distribution, reconstructed energy, and the reconstructed event location for both the prompt and delayed IBD candidate pair for the signal and background. 
These PDFs were generated using the RAT-PAC simulations by evaluating the detector response as described in Sections \ref{sec:signal} and \ref{sec:backgrounds}. 

The discussed sensitivity evaluations require that the confidence intervals are known for the two cases.
Simplified Monte Carlo experiments were generated 100,000 times from sampling the templates for Case 1 and 2 for each investigated dwell time. 
Each Monte Carlo simplified experiment was then fit using a negative log-likelihood approach to the corresponding template which was scaled to the appropriate detector dwell time by minimizing $\mathcal{L}_{IBD}$ determined from the prompt and delayed statistic, $\mathcal{L}_{p}$ and $\mathcal{L}_{d}$. 
Here, $\mathcal{L}$ refers to the log-likelihood fit performed using ROOFIT \cite{roofit} and MINUIT \cite{minuit}.
\begin{equation}
    \mathcal{L}_{IBD}= \mathcal{L}_{p}+\mathcal{L}_{d}.
\end{equation}
In this study bin-to-bin correlations have not yet been taken into account, and the covariance matrix is assumed to be diagonal.

\begin{figure}
\includegraphics[width=0.98\linewidth]{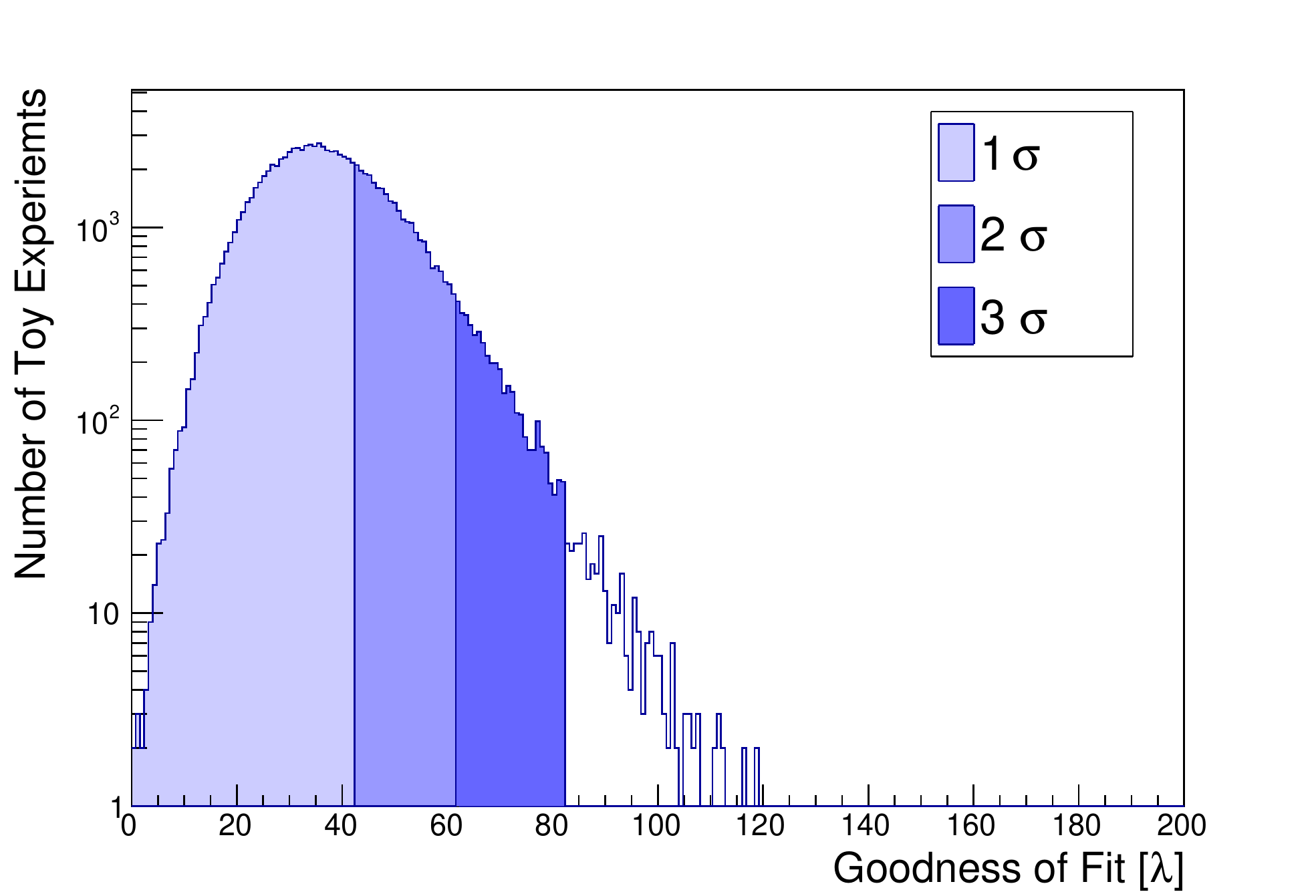}

\caption{ Defined confidence interval based on the test statistic $\lambda$ for the sensitivity to exclude experiments from the hypothesis that there are no nuclear reactors.}
\label{fig:confidence}.
\end{figure}

The sensitivity test statistic, $\lambda(\mu)$, is determined by comparing the result of the fit for experiment $\mu$ to the best fit result for another simplified Monte Carlo experiment $\mathcal{L}_{\mu_{best}}$:

\begin{equation}
    \lambda(\mu)=2(\mathcal{L}_{\mu}-\mathcal{L}_{\mu_{best}})
\end{equation}
Figure \ref{fig:confidence} demonstrates the test statistic, $\lambda$, for the Case 1 analysis.
A histogram of the resulting test statistics is generated, and the confidence intervals (1-3$\sigma$) can be defined from the sample based on the spread relative to the minimum value of $\lambda(\mu)$.

Similar to the process used to define the confidence intervals, a frequentest approach was used to generate test statistics for each of the experiments to compare the standard template of Case 1 $\&$ 2 \cite{feldman1998unified}. 
Again, 100,000 simplified Monte Carlo experiments were run for each detector standoff, dwell time, and reactor power. 
Following the results of the fits, the distribution of the critical values were recorded. 
For each distribution a delineation is made to determine where 90$\%$ of the experiments fit. 
This value, $\lambda_{90}$, is used in combination with the previously determined confidence intervals for Case 1 $\&$ 2 to determine the confidence level to which an experiment agrees with a prior assumption.

\section{Results \label{sec:results}}
\begin{figure*}[t]

\includegraphics[width=0.325\linewidth]{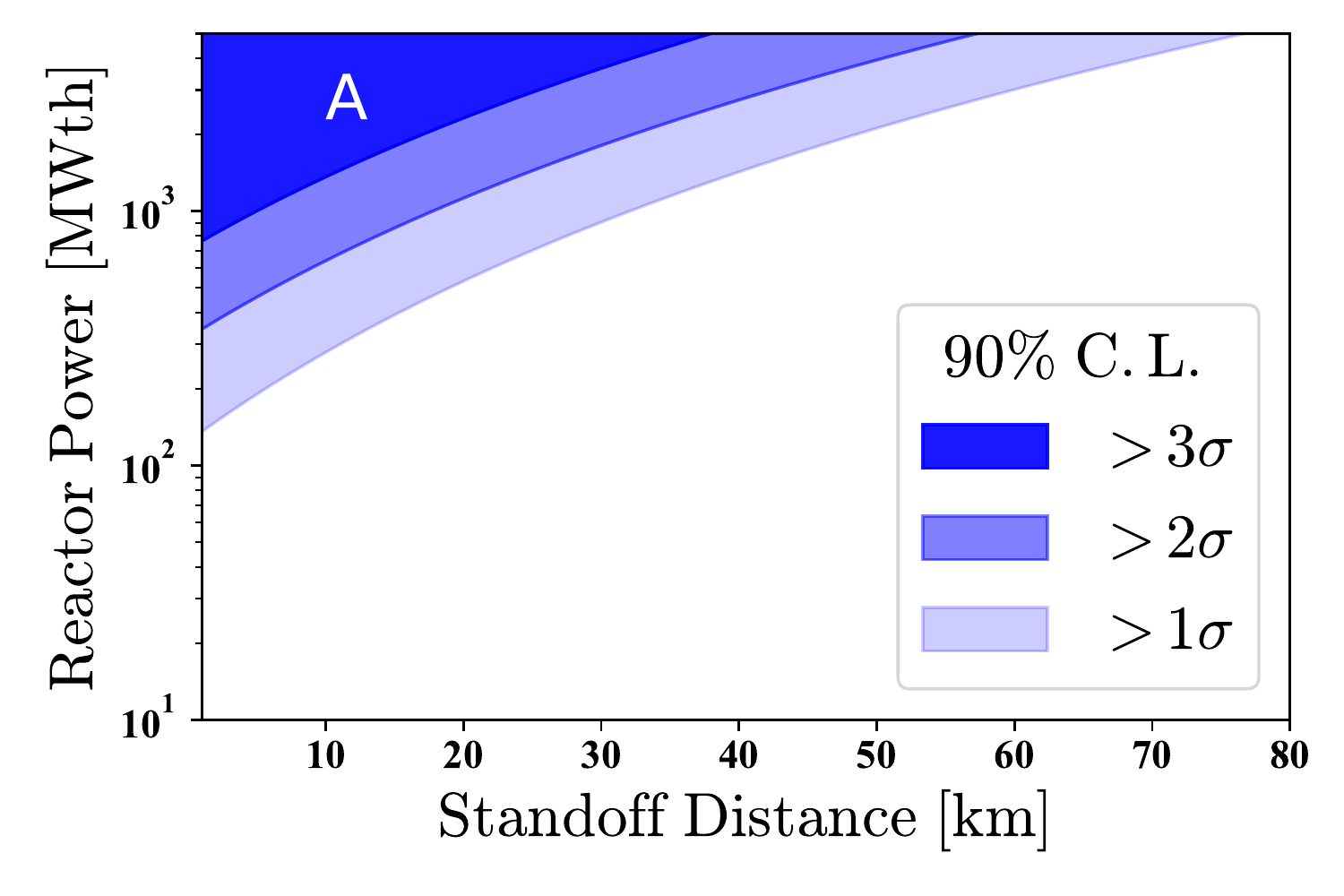}\includegraphics[width=0.325\linewidth]{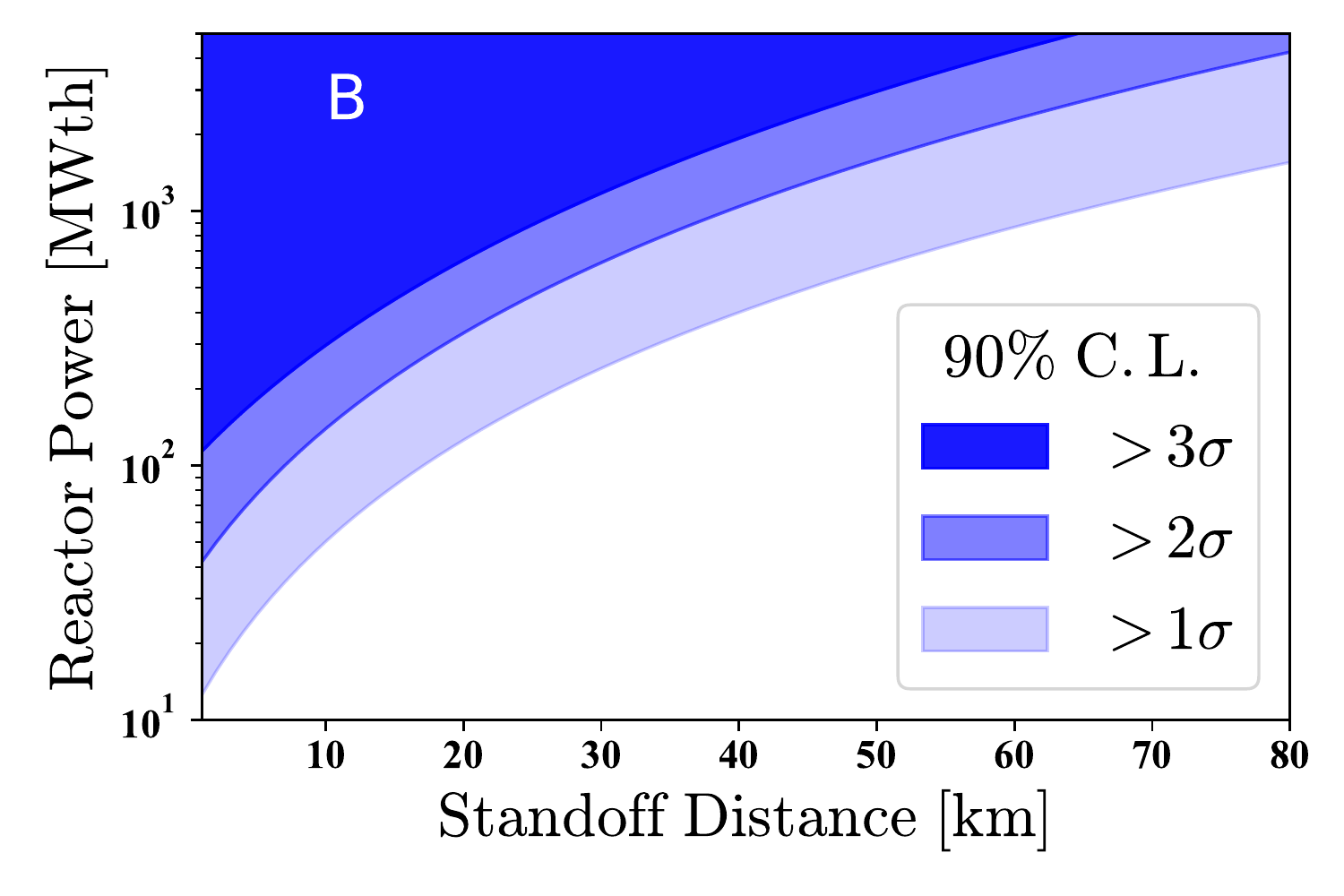}\includegraphics[width=0.325\linewidth]{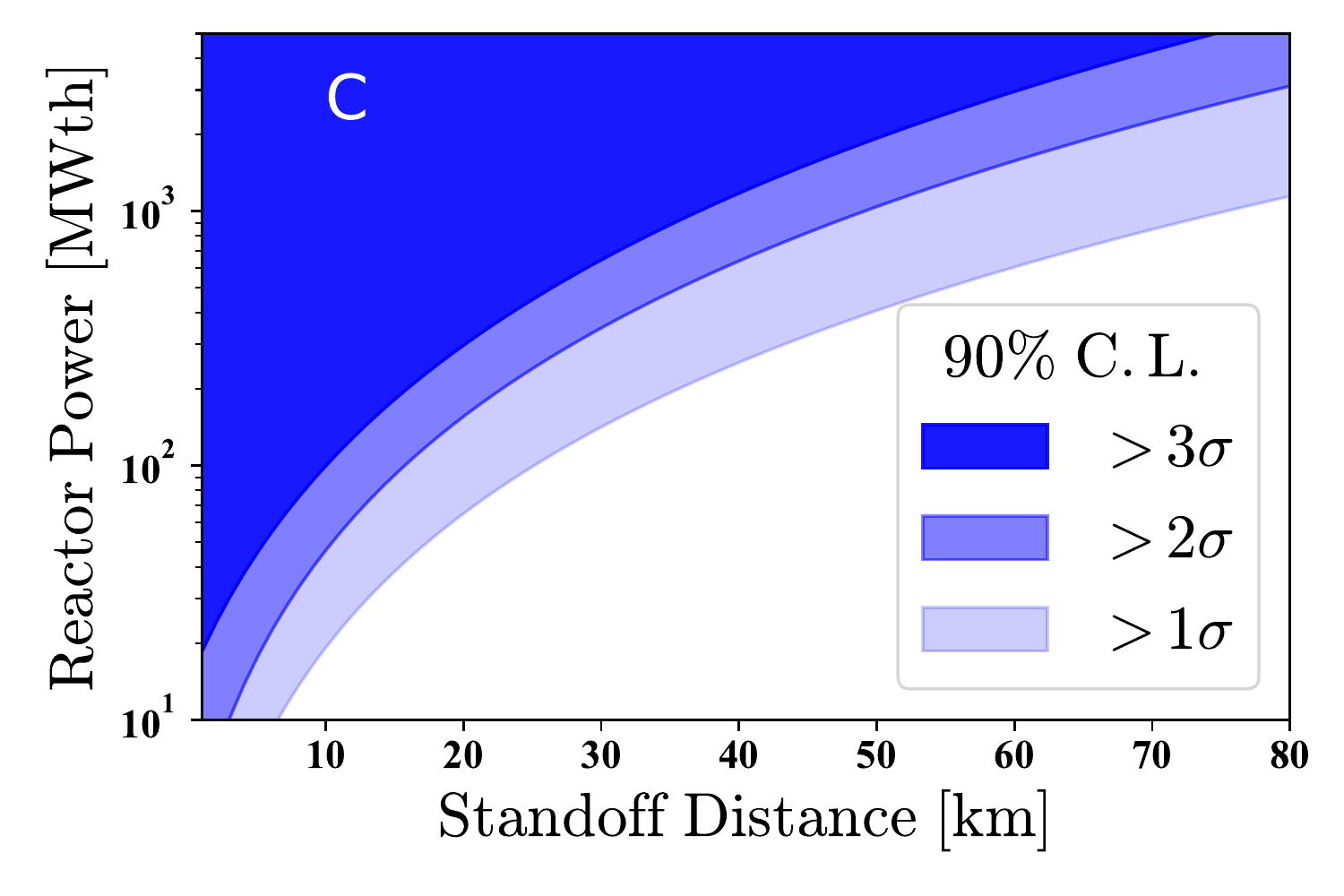}

\caption{ Sensitivity of the WATCHMAN detector to exclude or discover an undeclared nuclear reactor as a function of thermal power and standoff distance for a dwell time of one month [A], three months [B], and one year [C]. Experiments inside of the contour bands indicate scenarios in which the reactor power and standoff would disagree with the hypothesis that there are no nuclear reactors present with a 90 $\%$ confidence level. The confidence interval of 90$\%$ represents the percentage of simplified Monte Carlo for a given reactor power and standoff fall within the 1, 2, and 3 $\sigma$ determined by statistical tests of the simplified Monte Carlos of the no reactor hypothesis against the associated PDF template. }
\label{fig:case1}.
\end{figure*}

\begin{figure*}[t]

\includegraphics[width=0.325\linewidth]{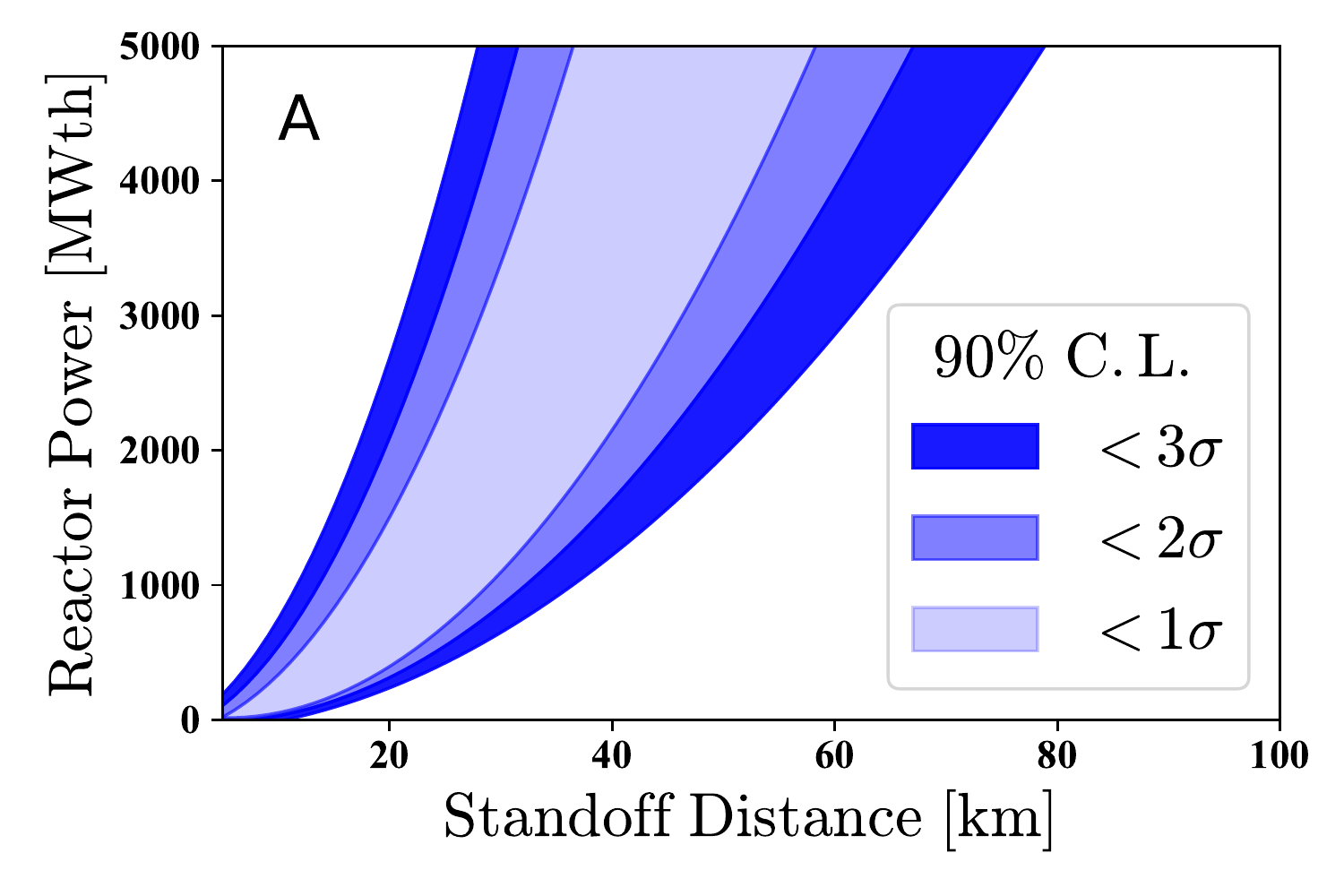}\includegraphics[width=0.325\linewidth]{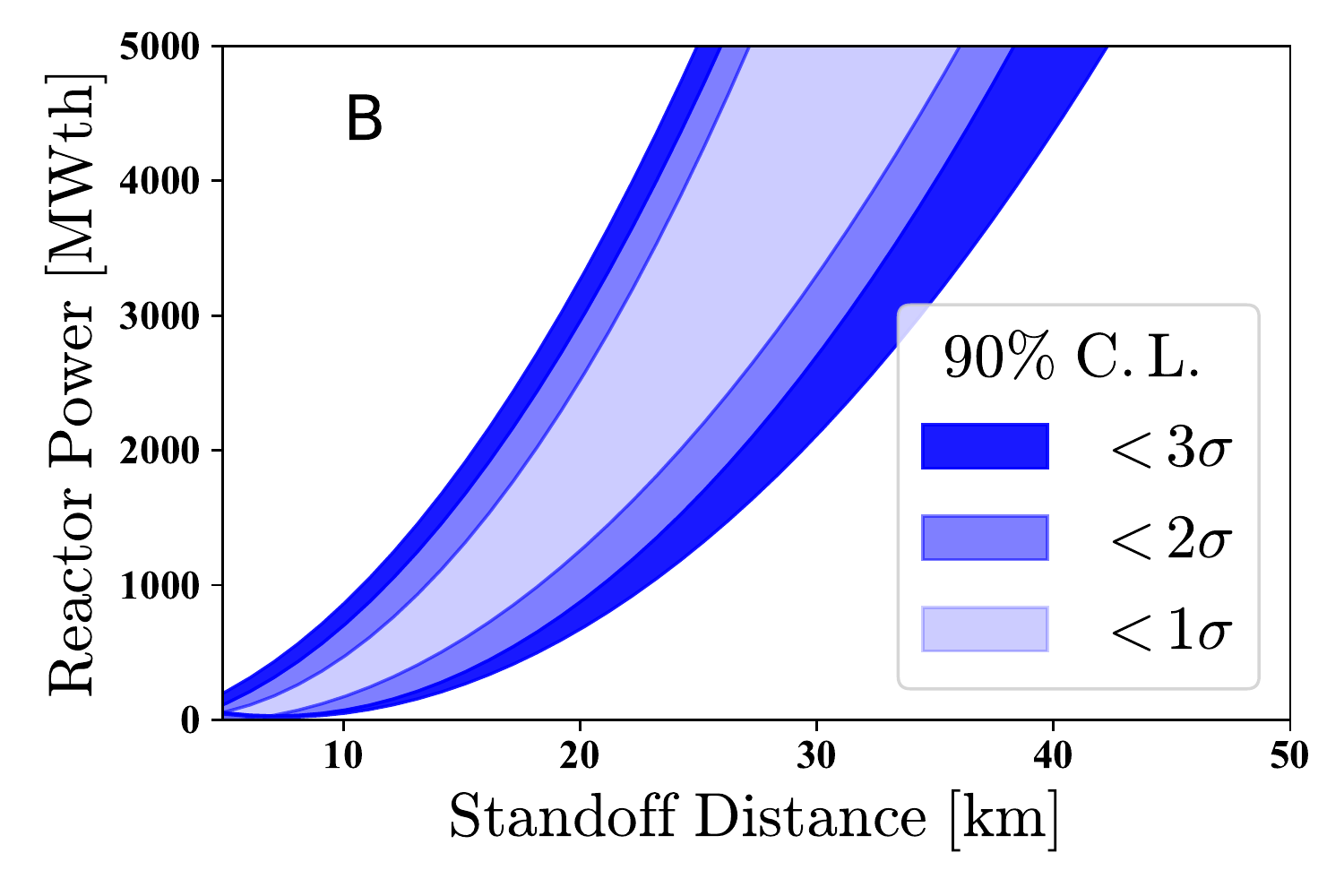}\includegraphics[width=0.325\linewidth]{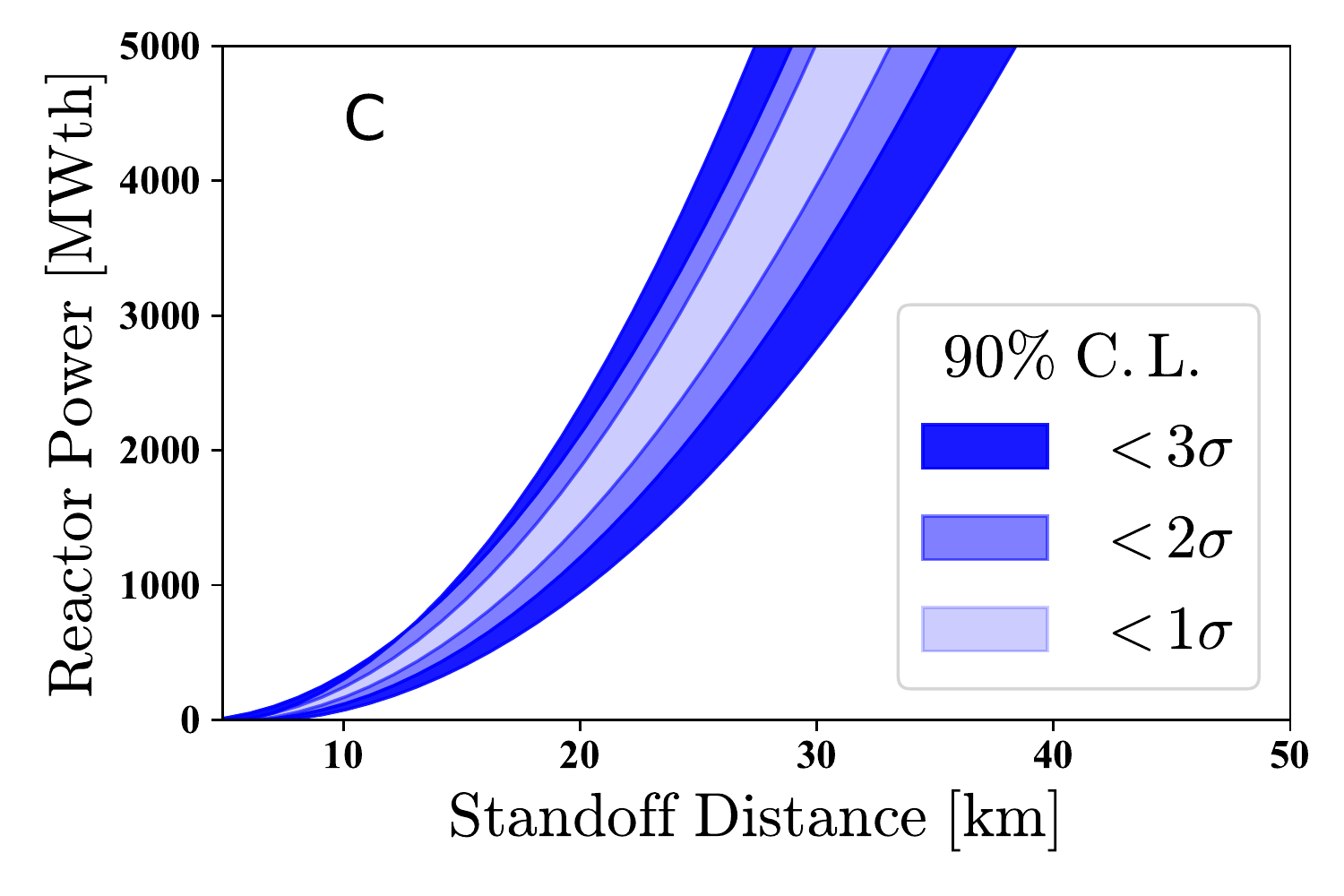}

\caption{ Sensitivity of the WATCHMAN detector to confirm the operating power of the Hartlepool Reactor complex for a dwell time of one month [A], three months [B], and one year [C]. Experiments outside of the contour bands indicate scenarios in which the reactor power and standoff would disagree with the hypothesis that there is a 3GW\textsubscript{th} reactor at a 26 km standoff greater than 3$\sigma$ with a 90 $\%$ confidence level. The confidence interval of 90$\%$ represents the percentage of simplified Monte Carlo for a given reactor power and standoff fall within the 1, 2, and 3 $\sigma$ determined by statistical tests of the simplified Monte Carlos of a reactor complex with a 26 km standoff with 3 GW\textsubscript{th} hypothesis against the associated PDF template.}
\label{fig:case2}.
\end{figure*}

Current safeguards documents used by the IAEA do not explicitly mention the timeliness goals for an entire nuclear reactor facility, but rather for significant quantities of material. 
A hidden nuclear reactor can contain or produce direct use material, irradiated nuclear material, and uranium with less than 20 $\%$ concentration of \textsuperscript{235}U \cite{glossary2002edition}.
In these three cases the timeliness goal for the discovery of a significant quantity is one month, three months, and one year, respectively; for which the significant quantities are 8 kg of plutonium or 75 kg of low enriched uranium.  
The sensitivity of Case 1 and Case 2 was simulated for these three dwell times to evaluate the ability to exclude unknown nuclear reactors from that region. 

Consistent with IAEA guidelines, the minimum reactor power capable of yielding the diversion of a significant quantity of plutonium 
is approximately 50 MW\textsubscript{th} \cite{national2005monitoring}. 
The 50 MW\textsubscript{th} is based on the expectation that the amount of plutonium, $M_{Pu}$ (in grams), that can be generated in a nuclear reactor scales with the thermal power, $P_{th}$, by:
\begin{equation}
    M_{Pu} \approx 	0.25\left[\dfrac{g}{MW_{d}}\right]\times P_{th}.
\end{equation}

Additionally, both typical commercial nuclear reactors and industrial plutonium generation reactors operate with a thermal powers on the order of GW\textsubscript{th} . As a result we evaluate the bounds of these thermal powers to better generalize the use cases of far-field reactor monitoring with a kiloton scale water Cherenkov detector. 

\subsection{\label{sec:EandDreactors}Case 1: Exclusion and Discovery}
In the application of a kiloton-scale water-Cherenkov antineutrino detector to exclude the existence of nuclear reactors from a region, one must compare the expected signal if no reactors are present to the detected antineutrino signal.
 Figure \ref{fig:case1} shows exclusion contours for the existence of a nuclear reactor in a region around the Boulby mine location (assuming the Hartlepool reactor complex does not exist) over dwell times of one month, three months, and one year. The reactor standoff and power consistent with no reactors present are shown as being in agreement to one, two, and three $\sigma$ to a 90$\%$ confidence limit.

Given the range of the reactor standoffs, the technology at this scale is unlikely to be used to covertly monitor or discover nuclear reactors across country borders. 
A sensitivity to discover nuclear reactors on the scale of 10s of km requires that there is a host country within that distance for the suspected reactor location willing to deploy such a detector.
However, the results demonstrate the capability to confirm the absence of undeclared nuclear reactors under a possible cooperative verification regime. 
For example, we consider a possible scenario in which a host facility previously had a nuclear reactor at a specific site. Despite the shutdown of nuclear reactor operations, the facility is still used as a laboratory to support other defense science or technologies that the host country intends to keep covert, without violating the terms of the verification regime. 
The results shown in this section highlight the level of non-intrusiveness (standoff) that is available based on the nuclear reactor power. 
Additionally, this technology can be used as a deterrent in a cooperative verification scenario. 
Locations and regions where reactors would be difficult for surveillance and other activities can adopt this technology to verify the absence of nuclear reactors.

Nuclear reactors with power outputs of 10's of MW\textsubscript{th} can be observed above background to 3 $\sigma$ with 90$\%$ confidence at a range of no more than 10 km for a one year dwell time.
Given this restricted range, the antineutrino facility used to monitor nuclear reactors on the scale of research reactors will likely need to be within or near the actual facility that a previously existing nuclear reactor was located. 
Depending on the required limit set as an anomaly trigger, events that indicate the presence of a nuclear reactor are experiments that the Gd-H$_2$O baseline design would be able to observe above background. 
Specifically, a 50 MW\textsubscript{th} nuclear facility can be excluded or found from a 5 km standoff within one year, while the Hartlepool nuclear reactor complex is seen above background within the first month.

Similarly, nuclear reactors on order of 100 MW\textsubscript{th} can be excluded up to 30 km distance within one year. This allows for less intrusive monitoring of an entire geographical region.
Nuclear reactors on order of GW$_{th}$ are traditionally associated with commercial nuclear reactors; however, industrial plutonium production reactors have historically had similar thermal powers \cite{diakov2011history}. 

The results presented here  generalize the application of this detector design to a range of standoff distances and reactor power levels. In the hypothetical deployment configuration, the experiment will monitor two 1500 MW\textsubscript{th} nuclear reactors at a 26 km standoff. 
Based on the results shown in Figure \ref{fig:case1}, the  baseline Gd-H$_2$O detector design will be able to identify the presence of the Hartlepool nuclear complex within a one month dwell time at greater than 3$\sigma$. Additionally, if one nuclear reactor were to shut down for maintenance or refueling for one month, the detector would be able to identify that there is a nuclear reactor above background with 3$\sigma$ confidence.  
Given the range of this verification technology, three sigma at 90$\%$ confidence is considered conservative.
In the case where there are other verification and monitoring tools in place, a lower confidence may be used to initiate a response.

\subsection{\label{sec:verificationofreactorpower}Case 2: Verification of Reactor Powers}
We perform the same analysis with the hypothesis of  a 3-GW\textsubscript{th} reactor power at a 26-km standoff to determine the sensitivity to both reactor ranging and reactor power confirmation.
The results of the capabilities to find a reactor in this context are shown in Figure \ref{fig:case2}. 
It is expected that distinct signatures in the oscillation pattern would allow the sensitivity to delineate between large thermal capacity reactors at a far standoff, and low-powered reactors at a close standoff. 
However, the combination of relatively low statistics, the energy resolution of positrons from Cherenkov light, and the energy threshold for the positron signal removes the signatures of the oscillation patterns.
To effectively determine the range of the reactor complex requires \textit{a priori} knowledge of the reactor power and vice versa. 

Despite the inability to delineate between reactor power and standoff, there are useful applications that can be identified from these capabilities. A nuclear reactor facility may have commercial reactors that are used to produce power. 
To share resources and personnel, another covert nuclear reactor specifically for military applications may also be on the same site. 
If this reactor is undeclared, it would not be included in a verification plan. Additionally, this technology may be applied to a bilateral treaty with an emphasis on reducing fissile material production. If the terms of such agreements were to limit the production of plutonium, a Gd-H$_2$O detector could measure the produced amount of plutonium generated without the intrusive step of being in close proximity to the reactor.

This hypothetical deployment was intended to monitor a facility with a total capacity of 3 GW\textsubscript{th}. With a one year dwell time, if the Gd-H$_2$O design were used, the power of the Hartlepool reactors  could be constrained to within 500 MW\textsubscript{th} at 3$\sigma$. Additionally, since the Hartlepool complex has two 1500 MW\textsubscript{th} nuclear reactors, an attempt will be made to detect one nuclear reactor in the presence of the other.

Based on the results in Figure \ref{fig:case2}, the  detector  will be capable of observing that both reactors are operating within four months at 3$\sigma$ with 90$\%$ confidence. 
One concern about this approach is that it is susceptible to reactor operator malfeasance.
In the case that the host facility were attempting to mask a small reactor signal, they could simply report a higher thermal output of the existing reactors.

\section{Summary and Outlook \label{sec:summary}}
In this study we have used the WATCHMAN baseline Gd-H$_2$O detector design to determine the capability of such a detector to exclude the presence of an undeclared nuclear reactor, and to verify the thermal power and standoff of known reactor complexes. 
This provides context for the application for a nominal 1-kton scale detector for reactor monitoring.
To achieve these goals a detailed understanding of the detector response to all of the expected backgrounds, from accidentals stemming from radiation from internal detector components, muon induced fast neutrons and radionuclides, and the world antineutrino background was achieved through simulations. 

The world antineutrino background will be the largest background in this hypothetical deployment. 
Although sensitivity was identified for a range of reactor powers and standoffs, the application of this technology in another location on Earth would yield different results. 
The United Kingdom and surrounding countries have a large installed nuclear capacity, making the resulting antineutrino background a higher background relative to locations with fewer reactors. Therefore, claims about the sensitivity of this design are likely conservative compared to use of the same design in other geographical locations with lower reactor backgrounds. 
Conversely, for specific use cases and greater standoffs, larger fiducial volumes may be necessary. Based on the recommendations from the NuTools study, it is unlikely that a larger-scale detector will have utility based on cost and excavation constraints \cite{akindele2021nu}.

The WATCHMAN baseline Gd-H$_2$O design highlights the application of known and proven technology previously used to measure solar neutrinos, antineutrinos from a supernova, and probe the possibility of proton decay, to aid in nonproliferation efforts to discover nuclear reactors or verify their operation \cite{sno_solar,fukuda2003super}. Further progress in reactor discovery sensitivity can be explored through the use of photon detection enhancements such as wavelength shifting plates and Winston cones, through scalable media that produce more light from interactions such as water-based liquid scintillator, and antineutrino directionality. All of such measures are currently under investigation by the WATCHMAN collaboration\cite{fukuda2003super,adams2001clas,yeh2011new,land2020mev,ford, mullen}. 

\acknowledgements
We thank the WATCHMAN collaboration for the review of this article and helpful comments. This material is based upon work supported by the Department of Energy National Nuclear Security Administration under Award Number DE-NA0000979 through the Nuclear Science and Security Consortium, and under Award Number DE-NA0003920 through the Monitoring, Technology, and Verification Consortium. This work was performed under the auspices of the U.S. Department of Energy by Lawrence Livermore National Laboratory under Contract DE-AC52-07NA27344 and released through LLNL-JRNL-814121. 
This work was also sponsored through the Scientific and Technology Facilities Council.
We thank Jason Brodsky at LLNL for useful discussions and guidance when evaluating sensitivity methodologies; and Michael Mendenhall and Ramona Vogt for guidance navigating FREYA with our detector response. 
 

\bibliographystyle{apsrev4-1}
\bibliography{wmsen.bib}


\end{document}